\documentclass{statsoc}
\usepackage{amsmath,amssymb}
\usepackage{graphicx}
\usepackage{upgreek}

\numberwithin{equation}{section}
\newtheorem{thm}{\textit{Theorem}}[section]

\newtheorem{cond}{\textit{Condition}}[section]

\title[Comparing time-varying spectra based on A-D statistic]{Tests for comparing time-invariant and time-varying spectra based on the Anderson-Darling statistic}
\author[S. Zhang]{Shibin Zhang (Corresponding author)}
\address{Department of Mathematics, Shanghai Normal University, 100 Guilin Rd., Shanghai 200234, China.}
\email{zhang\_shibin@shnu.edu.cn}
\author[S. Zhang {\it et al.}]{Xin M. Tu}
\address{Division of Biostatistics and Bioinformatics, Department of Family Medicine and Public Health, UC San Diego, La Jolla, California 92093, USA.}
\email{x2tu@ucsd.edu}

\begin{document}

\begin{abstract}
Based on periodogram-ratios of two univariate time series at different frequency points, two tests are proposed for comparing their spectra. One is an Anderson-Darling-like statistic for testing the equality of two time-invariant spectra. The other is the maximum of Anderson-Darling-like statistics for testing the equality of two time-varying spectra. Both of two tests are applicable for independent or dependent time series. Several simulation examples show that the proposed statistics outperform those that are also based on periodogram-ratios but constructed by the Pearson-like statistics.
\end{abstract}

\section{Introduction}
\label{sec:intro}

Comparison of spectra has wide applications in many fields of research and practice. Research in statistical methodology for comparing spectra has been attracting considerable interest for several decades.

Most existing literature on comparison of spectra assume that spectra are time-invariant, e.g.~\cite{Diggleetal1991}, \cite{Detteetal2011}, \cite{DecowskiLi2015}, \cite{Eichler2008} and \cite{JentschPauly2015}. By transforming the problem to the goodness-of-fit (gof) test that the periodogram-ratios of two univariate stationary time series at different frequencies are sampled from $F(2,2)$, \cite{ZhangTu2018} proposed a Pearson-like statistic to test the equality of two time-invariant spectra, where $F(2,2)$ denotes an $F$ distribution with 2 and 2 degrees of freedom. Using blocking, they further extended the approach to the more general setting of comparing two time-varying spectra of locally stationary time series.

A major limitation of the gof test based on the Pearson statistic is its dependence on the partition of the sample space. In this paper, we propose a new approach to address this key limitation. The proposed approach is based on Anderson-Darling (A-D) statistics of the form
\begin{equation}
  \hat{A}_n=\hat{A}_n(x_1,\cdots,x_n)=n\int_0^\infty \frac{(\hat{F}_n (x)-F(x))^2}{F(x)(1-F(x))}\,\mathrm{d}F(x), \label{eqn:andersondarlingstatistic}
\end{equation}
where $F(x)$ is the cumulative distribution function (cdf), and $\hat{F}_n (x)$ is the empirical cdf of observations $x_1,\cdots,x_n$. In our setting of comparing two spectra, the observations are periodogram-ratios of two univariate time series at different frequencies. Our proposed statistics have some advantages. First, the proposed statistics are easy to compute and invariant with respect to the periodogram-ratios, i.e., the value of the the statistic is unchanged after exchanging their places in the numerator and denominator. Second, the test is quite powerful. In the paper, the test for comparing spectra becomes the gof test that the periodogram-ratios are sampled from the $F(2,2)$. Since the $F(2,2)$ is heavy-tailed and has the reciprocal exchangeability, it is quite sensitive to detecting differences in the tails. It makes the A-D statistic quite powerful to reject the null since differences occur in the tails~\citep{AndersonDarling1954}. Our simulation study also confirms the second feature of the approach.

The rest of this article is organized as follows. Sections~\ref{sec:comparingstationaryspectra} and \ref{sec:comparingtimevaryingspectra} present the proposed statistics and their asymptotic properties for comparing time-invariant as well as time-varying spectra. Section~\ref{sec:numexample} reports results for examining performance of the proposed tests using simulated data. Section~\ref{sec:conclusion} contains our concluding remarks. All proofs are relegated to the Appendix. Throughout the paper, $|a|$, $\overline{a}$, $\Re(a)$ and $\Im(a)$ denote the complex modulus, conjugate, real part and imaginary part of a complex number $a$, respectively. For matrix notation, $A^\tau$ denotes the transpose of a matrix $A$, $A^\ast$ denotes the conjugate transpose of a complex-valued matrix $A$, and $\|A\|$ denotes the Euclidean norm of $A$. For ease of notation, we write $C$ for any generic positive constant.

Supplementary materials related to this article, including some R programs and a guide for using them, are available online.

\section{Comparing time-invariant spectra}
\label{sec:comparingstationaryspectra}

\subsection{Model}

Let $\{\mathbf{x}_t, t\in \mathbb{Z}\}$ denote a bivariate stationary process with values in $\mathbb{R}^2$ which has a linear representation of the form
\begin{equation}
  \mathbf{x}_t=(x_{1,t},x_{2,t})^\tau=\sum_{n=-\infty}^\infty \Psi_n Z_{t-n}, \quad \textrm{where }\sum_{n=-\infty}^{\infty}\|\Psi_n\| |n|^{1/2}<\infty, \label{eqn:ageneralizedlinearprocess}
\end{equation}
and $\{Z_t,t\in \mathbb{Z}\}$ is a sequence of independent identically-distributed (i.i.d.) random variables, with mean $\mathbf{0}$, covariance matrix $\Sigma$ and finite fourth moment.

We assume that we have a sufficiently large number, $T$, of observations from the bivariate zero-mean stationary time series in (\ref{eqn:ageneralizedlinearprocess}), $\mathbf{x}_t$. The spectral density matrix is given by
\begin{equation}
  \mathbf{f}(\omega)=\big(f_{ij}(\omega)\big)_{2\times 2}=(2\pi)^{-1}\mathcal{A}(\omega)\Sigma \mathcal{A}^\ast(\omega) \label{eqn:fomegadetail}
\end{equation}
for $\omega\in [-\pi,\pi]$, where $\mathcal{A}(\omega)=(
  A_{ij}(\omega)
)_{2\times 2}=\sum_{n=-\infty}^\infty \Psi_n \mathrm{e}^{-\mathrm{i} n \omega}$~\citep[by e.g.][Chapter~4]{ShumwayStoffer2011}.
From (\ref{eqn:fomegadetail}), we know that the spectral density $\mathbf{f}(\omega)$ is continuous with respect to $\omega$.
The discrete Fourier transform (DFT) of the observation $\{\mathbf{x}_t,t=1,\cdots,T\}$ is
\begin{equation*}
  \mathbf{y}_T(\omega_k)=\frac{1}{\sqrt{2\pi T}}\sum_{t=1}^T \mathbf{x}_t \mathrm{e}^{\mathrm{i} \omega_k t},
\end{equation*}
where $\omega_k=2\pi k/T$, $k=0,1,\cdots,[T/2]$. We extend the definition of $\mathbf{y}(\cdot)$ to a piecewise constant function on $[-\pi,\pi]$ as follows:
\begin{equation}
  \mathbf{y}_T(\omega)=\begin{cases}
    \mathbf{y}_T(\omega_k), & \textrm{if }\omega_k-\pi/T<\omega\leq \omega_k+\pi/T \textrm{ and }0\leq \omega\leq \pi,\\
   \mathbf{y}_T(-\omega), & \textrm{if } \omega\in [-\pi,0).
  \end{cases} \label{eqn:extentionmathbfy}
\end{equation}

For any frequency $\omega\in [-\pi,\pi]$, we define the periodogram by \begin{equation*}
\mathbf{I}_T(\omega)=\big( I_{T,ij}(\omega)\big)_{2\times 2}=\mathbf{y}_T(\omega) \mathbf{y}_T^\ast(\omega).
\end{equation*}
Then the periodogram is also a piecewise constant function on $[-\pi,\pi]$, in accordance with the definition of \cite{Fuller1996}.

\subsection{The test statistic}

In this paper, we test the null:
\begin{equation}
  H_0:\ f_{11}(\omega)=f_{22}(\omega) \quad \textrm{on } (0,\pi). \label{eqn:nullhypothesis_equalityofspectraldensities}
\end{equation}

The idea for the construction of the test statistic is originated from periodogram properties of the independent time series. For each $i=1,2$, if $f_{ii}(\omega)>0$ for all $\omega \in (0,\pi)$, then the random vector $\big(2I_{T,ii}(\omega_1)/f_{ii}(\omega_1),\cdots,2I_{T,ii}(\omega_m)/f_{ii}(\omega_m)\big)^\tau$ converges in distribution as $T\rightarrow \infty$ to a vector of i.i.d. $\chi^2(2)$ random variables for $0<\omega_1<\cdots<\omega_m<\pi$, where $\chi^2(2)$ denotes a chi-squared distribution with 2 degrees of freedom~\citep[e.g.][Theorem~10.3.2]{BrockwellDavis1991}.

When two components of $\mathbf{x}_t$, $x_{1,t}$ and $x_{2,t}$, are independent, $\{2I_{T,11}(\omega)/f_{11}(\omega),\,\omega\in (0,\pi)\}$ and $\{2I_{T,22}(\omega)/f_{22}(\omega),\,\omega\in (0,\pi)\}$ are also mutually independent. Then the test for the null (\ref{eqn:nullhypothesis_equalityofspectraldensities}) is transformed to the gof test
\begin{equation*}
  \frac{I_{T,11}(\omega_l)}{I_{T,22}(\omega_l)},\,l=1,\cdots,m, \quad \textrm{are sampled from }F(2,2).
\end{equation*}
However, the independence of $2I_{T,11}(\omega)/f_{11}(\omega)$ and $2I_{T,22}(\omega)/f_{22}(\omega)$ does not hold if the dependence between components of the time series is unknown. Since the DFT series $\{\mathbf{y}_T(\omega_l),\,l=1,\cdots,m\}$ is asymptotic independent among different $\omega_l$'s, the periodograms of two components of $\mathbf{x}_t$ at different frequencies are asymptotically independent, no matter two components of $\mathbf{x}_t$ are independent or not~\citep[][Lemma~2.1]{ZhangTu2018}.

If two spectral densities are continuous in $(0,\pi)$ and $L$ is taken greater, the periodograms $I_{T,11}(\frac{l-1/2}{L}\pi)$ and $I_{T,22}(\frac{l}{L}\pi)$ are expected to estimate the respective spectral densities $f_{11}(\cdot)$ and $f_{22}(\cdot)$ at the same frequency point. By Lemma~2.1 of \cite{ZhangTu2018}, under the null (\ref{eqn:nullhypothesis_equalityofspectraldensities}), the periodogram-ratios
\begin{equation*}
  I_{T,11}(\frac{l-1/2}{L}\pi)/I_{T,22}(\frac{l}{L}\pi),\quad l=1,\cdots, L-1
\end{equation*}
are expected to behave like i.i.d. $F(2,2)$ distribution for a large $L$ as $T$ tends to infinity. Then the test for the null (\ref{eqn:nullhypothesis_equalityofspectraldensities}) is transformed to the gof test
\begin{equation}
  \frac{I_{T,11}(\frac{l-1/2}{L}\pi)}{I_{T,22}(\frac{l}{L}\pi)},\, l=1,\cdots, L-1,\quad \textrm{are sampled from }F(2,2). \label{eqn:goodnessoffit}
\end{equation}
Based on this idea, \cite{ZhangTu2018} proposed a Pearson-like statistic, with a limiting chi-squared distribution, to test (\ref{eqn:nullhypothesis_equalityofspectraldensities}). However, as it relies on the partition of the sample space, different partitions may result in different test results.

The criterion proposed by \cite{AndersonDarling1952} covers a broad class gof test statistics that are constructed by a measure of discrepancy or ``distance'' between the cdf and its empirical counterpart. Compared to others such as those proposed by \cite{AndersonDarling1952}, the Anderson-Darling statistic (\ref{eqn:andersondarlingstatistic})
has two advantages when testing a sample drawn from $F(2,2)$. First, the statistic $\hat{A}_n$ is easy to compute and its value is unchanged if replacing the observations $x_1,\cdots,x_n$ by their reciprocals $1/x_1,\cdots,1/x_n$. When $F(x)=\frac{x}{1+x}\mathbb{I}_{(0,\infty)}(x)$ is the cdf of $F(2,2)$, by (2) of \cite{AndersonDarling1954}, we obtain that the simplified expression of $\hat{A}_n$ is
 \begin{equation}
   \hat{A}_n(x_1,\cdots,x_n)=-n-\frac{1}{n}\sum_{i=1}^{n} (2 i-1)\big[\log(x_{(i)})-\log(1+x_{(i)})-
   \log(1+x_{(n-i+1)})\big], \label{eqn:simplifiedADstatistic}
  \end{equation}
  where $x_{(1)}<\cdots<x_{(n)}$ is the ordered series of $x_i$, $i=1,\cdots,n$. Since $x_{(i)}=1/y_{(n-i+1)}$ holds if $y_i=1/x_i$, $i=1,\cdots,n$, we obtain from (\ref{eqn:simplifiedADstatistic}) that $\hat{A}_n(x_1,\cdots,x_n)=\hat{A}_n(1/x_1,\cdots,1/x_n)$ holds. This results in that the value of $\hat{A}_n$ is invariant after exchanging places of numerator and denominator of the periodogram-ratios when testing the gof (\ref{eqn:goodnessoffit}) by the the A-D statistic. Second, the A-D statistic is quite powerful in testing a distribution deviating the $F(2,2)$. Since the $F(2,2)$ is heavy-tailed and has the reciprocal exchangeability (i.e., if $X\sim F(2,2)$ then $X^{-1}\sim F(2,2)$), differences from it is prone to be detected in the tails. The A-D statistic is the one that is quite powerful to detect the differences in the tails~\citep{AndersonDarling1954}.

Due to the properties of statistic (\ref{eqn:andersondarlingstatistic}), we propose an Anderson-Darling-like statistic
\begin{equation}
  \hat{A}_{T,L}=(L-1)\int_0^\infty \frac{(\hat{F}_{T,L} (x)-F(x))^2}{F(x)(1-F(x))}\,\mathrm{d}F(x), \label{eqn:proposedadstatistic}
\end{equation}
to compare two time-invariant spectra, i.e. test (\ref{eqn:goodnessoffit}), and to avoid choosing the partition of the sample space, where $L>1$ is an integer,
\begin{equation}
  \hat{F}_{T,L} (x)=\frac{1}{L-1}\sum_{l=1}^{L-1} \mathbb{I}_{(0,x]}\Big(\frac{I_{T,11}(\frac{l-1/2}{L}\pi)}{I_{T,22}(\frac{l}{L}\pi)}\Big),
  \label{eqn:hatFTL}
\end{equation}
and $F(x)=\frac{x}{1+x}\mathbb{I}_{(0,\infty)}(x)$.

\subsection{Sampling distribution Under the null}

To obtain the asymptotic property of $\hat{A}_{T,L}$ under the null hypothesis (\ref{eqn:nullhypothesis_equalityofspectraldensities}), we need the following condition.

\begin{cond}{\rm
 As $T\rightarrow \infty$, $T/L^2=O(1)$ and $L \log T/T \rightarrow 0$ (For clarity, we omit the dependence of $L$ on $T$).} \label{cond:condition2}
\end{cond}

Then, we have the following theorem.

\begin{thm}{\rm
Suppose that $f_{ii}(\omega)$, $i=1,2$, are bounded below away from zero and Lipschitz continuous, i.e., $f_{ii}(\omega)\geq \delta>0$ for all $\omega\in (0,\pi)$ and $i=1,2$, and $|f_{ii}(\omega_1)-f_{ii}(\omega_2)|\leq C |\omega_1-\omega_2|$, $i=1,2$,
hold for any $\omega_1\neq \omega_2$. Then, under Condition~\ref{cond:condition2}, if the null hypothesis (\ref{eqn:nullhypothesis_equalityofspectraldensities}) holds, we have
\begin{equation}
  \hat{A}_{T,L}\rightsquigarrow \int_0^1 \frac{B_0^2(t)}{t(1-t)}\,\mathrm{d}t,
  \label{eqn:mainresultofthm1}
\end{equation}
 where $B_0(t)$ is a standard Brownian bridge on $[0,1]$, and ``$\rightsquigarrow$'' denotes convergence in distribution.
} \label{thm:thm3.1}
\end{thm}

By Theorem~\ref{thm:thm3.1}, the statistic (\ref{eqn:proposedadstatistic}) can be used to test the null hypothesis (\ref{eqn:nullhypothesis_equalityofspectraldensities}). Given the pre-specified level of significance $\alpha$ ($0<\alpha<1$), the null hypothesis (\ref{eqn:nullhypothesis_equalityofspectraldensities}) is rejected if
\begin{equation}
  \hat{A}_{T,L}>a_{1-\alpha}, \label{eqn:testmethod_stationary}
\end{equation}
where $a_{1-\alpha}$ denotes the $1-\alpha$ quantile of the A-D test statistic with sample size tending to infinity. Under the null, the rejection probability of test (\ref{eqn:testmethod_stationary}) converges to $\alpha$ as $T$ goes to infinity. The cdf of the null distribution of the A-D statistic can be computed by using the algorithm of \cite{MarsagliaMarsaglia2004}. The quantiles can be computed by root-finding. The functions for computation can be found in the R package \texttt{goftest}~\citep{Farawayetal2015}.

\section{An extension for comparing time-varying spectra}
\label{sec:comparingtimevaryingspectra}

\subsection{Model}

We base our discussion on a multivariate extension~\citep[cf.][]{ZhangTu2018} of the model of \cite{Dettetal2011b}.

Let $\{\mathbf{x}_{t, T}, t=1,\cdots,T\}$ ($T\in \mathbb{N}$) be a bivariate locally stationary time series, where each observation $\mathbf{x}_{t, T}$ exhibits a linear representation of the form
\begin{equation}
  \mathbf{x}_{t,T}=\sum_{n=-\infty}^\infty \Psi_{t,T,n} Z_{t-n}, \quad t=1,\cdots, T.\label{eqn:xtT}
\end{equation}
We also assume that $\{Z_t,t\in \mathbb{Z}\}$ is a sequence of i.i.d. random variables, with mean $\mathbf{0}$, covariance matrix $\Sigma$ and finite fourth moment. We further assume the coefficients $\{\Psi_{t,T,n}\}$ behave like some smooth functions in a neighborhood of time $t/T$. Therefore we adopt not only the usual condition $\sum_{n=-\infty}^\infty \|\Psi_{t,T,n}\| <\infty$, but impose additionally that there exists a matrix-valued function $\Psi_n:\,[0,1]\rightarrow \mathbb{R}^{2\times 2}$ and a constant $C$ with
\begin{equation}
  \sum_{n=-\infty}^\infty \sup_{t=1,\cdots,T} \|\Psi_{t,T,n}-\Psi_n(t/T)\|\leq C\frac{1}{T}. \label{eqn:condnonstationary2}
\end{equation}
Furthermore, we assume that
\begin{equation}
  \sum_{n=-\infty}^\infty \sup_{u\in [0,1]}\|\Psi_n(u)\| |n|^{1/2}<\infty,  \label{eqn:cond.nonstationary1}
\end{equation}
and each element of $\Psi_n$ is a continuously differentiable function of $u$, and
\begin{equation}
  \sum_{n=-\infty}^\infty \sup_{i,j} \sup_{u\in[0,1]}|\mathrm{d}\Psi_n^{(i,j)}(u)/\mathrm{d}u|<\infty. \label{eqn:condnonstationary3}
\end{equation}
\cite{ZhangTu2018} gave a detailed discussion on the conditions (\ref{eqn:condnonstationary2})-(\ref{eqn:condnonstationary3}).
The time-varying spectral density of the locally stationary process $\{x_{t,T}\}$ is defined in terms of auxiliary function $\Psi_n$~\citep[cf.][]{Dettetal2011b}, that is,
\begin{equation}
  \mathbf{f}(u,\omega)=(
    f_{ij}(u,\omega)
  )_{2\times 2}=(2\pi)^{-1} \mathcal{A}(u,\omega) \Sigma \mathcal{A}^\ast(u,\omega), \quad u\in [0,1],\ \omega\in [-\pi,\pi] \label{eqn:timevaryingspectrum}
\end{equation}
where $\mathcal{A}(u,\omega)=\sum_{n=-\infty}^\infty \Psi_n(u)\mathrm{e}^{-\mathrm{i}n \omega}$.

Assume without loss of generality that the total sample size $T$ can be decomposed as $T=MB$, where $B$ is an integer and $M$ is an even integer. The main idea is to split the entire data into $B$ blocks with $M$ observations each, from which we define appropriate local periodograms. Specifically, let
\begin{equation}
  \mathbf{I}_{M}(u,\omega_k):=( I_{M,ij}(u,\omega_k))_{2\times 2}=\mathbf{y}_M(u,\omega_k)\mathbf{y}_M^\ast(u,\omega_k) \label{eqn:mathbfITtomegak}
\end{equation}
be the usual periodogram around $u$ computed from $M$ observations, that is, we set
\begin{equation}
  \mathbf{y}_M(u,\omega_k)=\frac{1}{\sqrt{2\pi M}}\sum_{s=1}^M \mathbf{x}_{[uT]-M/2+s, T}\mathrm{e}^{\mathrm{i}\omega_k s} \label{eqn:yMuomegak}
\end{equation}
and $\mathbf{x}_{t,T}=\mathbf{0}$, if $t\notin\{1,\cdots,T\}$ \citep{Dahlhaus1997}, where $\omega_k=2\pi k/M$, $k=0,1,\cdots,[M/2]$. For each $u$, we extend the definition of $\mathbf{y}_M(u,\cdot)$ to a piecewise constant function on $[0,1]\times[-\pi,\pi]$, the same extension method as in (\ref{eqn:extentionmathbfy}). Accordingly, we extend the definition of $\mathbf{I}_{M}(\cdot,\cdot)$ in (\ref{eqn:mathbfITtomegak}) to the periodograms on $[0,1]\times[-\pi,\pi]$.

\subsection{The test statistic}

In this section, we consider the problem of testing
\begin{equation}
  H_0:\ f_{11}(u,\omega)=f_{22}(u,\omega)\quad \textrm{on } \{\omega:\omega\in (0,\pi)\}\quad \textrm{for each }u\in[0,1]. \label{eqn:nullhypothesis_equalityofspectraldensitiesnonstationary}
\end{equation}

We use the notation $u_k=\frac{(k-1) M+M/2}{T}$ ($k=1,\cdots,B$) for the mid-point of each block, and for each integer $L>1$, define
\begin{equation}
  \hat{F}_{M,L}^{(k)} (x)=\frac{1}{L-1}\sum_{l=1}^{L-1} \mathbb{I}_{(0,x]}\Big(\frac{I_{M,11}(u_k,\frac{l-1/2}{L}\pi)}{I_{M,22}(u_k,\frac{l}{L}\pi)}\Big).
  \label{eqn:hatFTLeachblock}
\end{equation}
Let
\begin{equation}
  \hat{A}_{M,L}^{(k)}=(L-1)\int_0^\infty \frac{(\hat{F}_{M,L}^{(k)} (x)-F(x))^2}{F(x)(1-F(x))}\,\mathrm{d}F(x), \label{eqn:proposedadstatisticeachblock}
\end{equation}
where $\hat{F}_{M,L}^{(k)} (x)$ is defined by (\ref{eqn:hatFTLeachblock}), and $F(x)$ is defined as in (\ref{eqn:hatFTL}). In the $k$-th block, we use (\ref{eqn:proposedadstatisticeachblock}) to compare two local spectra. Intuitively, we should reject (\ref{eqn:nullhypothesis_equalityofspectraldensitiesnonstationary}) when at least one of $\hat{A}_{M,L}^{(k)}$, $k=1,\cdots,B$, exceeds a pre-specified threshold. Therefore, we propose a maximum of local A-D statistics
\begin{equation}
 M_{B,M,L}= \max_{k=1,\cdots,B}\{\hat{A}_{M,L}^{(k)}\} \label{eqn:teststatisticofnonstationarycase}
\end{equation}
to test (\ref{eqn:nullhypothesis_equalityofspectraldensitiesnonstationary}).

\subsection{Sampling distribution under the null}

To study the asymptotic behavior of $M_{B,M,L}$ under the null, we need the following condition.

\begin{cond}{\rm
As $T\rightarrow \infty$, $M\rightarrow \infty$, $M^2/T=O(1)$, $M/L^2=O(1)$ and $L \log M/M \rightarrow 0$ (For clarity, we omit the dependence of $L$ and $M$ on $T$).}\label{cond:condition2nonstationary}
\end{cond}

With arguments similar to the proof of Theorem~3.1 of \cite{ZhangTu2018},
we obtain the asymptotic properties of (\ref{eqn:teststatisticofnonstationarycase}) under the null below.

\begin{thm}{\rm
Suppose that the locally stationary time series $\{\mathbf{x}_{t, T}, t=1,\cdots,T\}$ ($T\in \mathbb{N}$) satisfies conditions (\ref{eqn:condnonstationary2})-(\ref{eqn:condnonstationary3}), the diagonal elements of $\mathbf{f}(u,\omega)$ are bounded blow away from 0 for all $(u,\omega)\in [0,1]\times(0,\pi)$, and $f_{11}(u,\omega)$ and $f_{22}(u,\omega)$ are uniformly Lipschitz continuous with respect to $\omega$ for $u\in [0,1]$, i.e. $|f_{ii}(u,\omega_1)-f_{ii}(u,\omega_2)|\leq C |\omega_1-\omega_2|$, $i=1,2$,
hold for any $\omega_1\neq \omega_2$ and some positive constant $C$ independent of $u$ and $\omega$. Then, under Condition~\ref{cond:condition2nonstationary}, if the null hypothesis (\ref{eqn:nullhypothesis_equalityofspectraldensitiesnonstationary}) holds, we have
\begin{equation}
  M_{B,M,L}=\max_{k=1,\cdots,B} \{\breve{A}_{M,L}^{(k)}\}+o_p\big(\frac{\sqrt{L\log M}}{\sqrt{M}}\big), \label{eqn:MBML-asyresult}
\end{equation}
where $\breve{A}_{M,L}^{(k)}$, $k=1,\cdots,B$, are i.i.d. random variables for each fixed $B$. Moreover, for each fixed $k$, $\breve{A}_{M,L}^{(k)} \rightsquigarrow \int_0^1 \frac{B_0^2(t)}{t(1-t)}\,\mathrm{d}t$ as $T$ goes to infinity.}\label{thm:thm3.2nonstationary}
\end{thm}

We propose the statistic (\ref{eqn:teststatisticofnonstationarycase}) to test the null hypothesis (\ref{eqn:nullhypothesis_equalityofspectraldensitiesnonstationary}). According to Theorem~\ref{thm:thm3.2nonstationary}, the cdf of $M_{B,M,L}$ is approximately equal to that of $\max_{k=1,\cdots,B}\{A_k\}$ in distribution,
where $A_k\stackrel{d}{=}\int_0^1 \frac{B_0^2(t)}{t(1-t)}\,\mathrm{d}t$, $k=1,\cdots,B$, are i.i.d. random variables. The cdf of $\max_{k=1,\cdots,B}\{A_k\}$ is $F_M(x)=F_A(x)^B$, where $F_A(x)$ is the cdf function of $\int_0^1 \frac{B_0^2(t)}{t(1-t)}\,\mathrm{d}t$. For a pre-specified level of significance $\alpha$ ($0<\alpha<1$), the null hypothesis (\ref{eqn:nullhypothesis_equalityofspectraldensitiesnonstationary}) is rejected if
\begin{equation}
  M_{B,M,L}>\kappa_{1-\alpha}(B), \label{eqn:testmethod_nonstationary}
\end{equation}
where $\kappa_{1-\alpha}(B)$ denotes the $1-\alpha$ quantile of the distribution of $\max_{k=1,\cdots,B}\{A_k\}$. Under the null hypothesis, the rejection probability of test (\ref{eqn:testmethod_nonstationary}) is approximately equal to $\alpha$.

The test (\ref{eqn:testmethod_nonstationary}) is convenient to use since $\kappa_{1-\alpha}(B)$ can be easily obtained by solving $F_A(x)^B=1-\alpha$.

\section{Numerical examples}
\label{sec:numexample}

In this section, we use simulations to evaluate small-sample performance of the proposed statistics (\ref{eqn:proposedadstatistic}) and (\ref{eqn:teststatisticofnonstationarycase}) when applied to test the null hypotheses (\ref{eqn:nullhypothesis_equalityofspectraldensities}) and (\ref{eqn:nullhypothesis_equalityofspectraldensitiesnonstationary}). All results are based on 1,000 Monte Carlo replications unless stated otherwise.

In all the simulation examples of Section~\ref{subsec:comp_stationaryspec}, we set $L=\min\{[T/4],[T^{3/4}]\}$. In the examples of Section~\ref{subsec:comp_timevaryingspec}, we set $B=[\sqrt{T}/5]$, $M=[T/B]$, and $L=\min\{[M/4],[M^{3/4}]\}$. These settings all satisfy Condition~\ref{cond:condition2} or \ref{cond:condition2nonstationary}.

\subsection{Comparing time-invariant spectra}
\label{subsec:comp_stationaryspec}

To show the performance of test (\ref{eqn:testmethod_nonstationary}) for Gaussian models, we consider nine stationary time series $\mathbf{x}_t=(
  x_{1,t},x_{2,t}
)^\tau$, $t=1,\cdots,T$ in this section. Their components are given as follows.

Model A (Copied $MA(1)$ model.) $x_{i,t}=Z_{i,t}-0.8 Z_{i,t-1}$, $i=1,2$.

Model B (Copied $MA(2)$ model.) $x_{i,t}=Z_{i,t}-0.8 Z_{i,t-1}-0.5 Z_{i,t-2}$, $i=1,2$.

Model C (Copied $AR(1)$ model.) $x_{i,t}=0.5 x_{i,t-1}+\sqrt{0.75} Z_{i,t}$, $i=1,2$.

Model D (Copied $ARMA(1,1)$ model.) $x_{i,t}=0.5 x_{i,t-1}+Z_{i,t}-0.5Z_{i,t-1}$, $i=1,2$.

Model E (Copied $AR(2)$ model.) $x_{i,t}=0.5 x_{i,t-1}-0.5 x_{i,t-2}+(1/\sqrt{1.5})Z_{i,t}$, $i=1,2$.

Model F ($MA(2)$-$MA(1)$ model.) $x_{1,t}=Z_{1,t}-0.8 Z_{1,t-1}-0.5 Z_{1,t-2}$ and $x_{2,t}=Z_{2,t}-0.8 Z_{2,t-1}$.

Model G ($AR(1)$-$ARMA(1,1)$ model.) $x_{1,t}=0.5 x_{1,t-1}+\sqrt{0.75} Z_{1,t}$ and $x_{2,t}=0.5 x_{2,t-1}+Z_{2,t}-0.5Z_{2,t-1}$.

Model H ($AR(1)$-$AR(2)$ model.) $x_{1,t}=0.5 x_{1,t-1}+\sqrt{0.75}Z_{1,t}$ and $x_{2,t}=0.5 x_{2,t-1}- 0.5 x_{2,t-2}+(1/\sqrt{1.5})Z_{2,t}$.

Model I ($AR(2)$-$AR(2)$ model.) $x_{1,t}=0.5 x_{1,t-1}-0.5x_{2,t-2}+(1/\sqrt{1.5})Z_{1,t}$ and $x_{2,t}=0.6 x_{2,t-1}-0.6 x_{2,t-2}+ \sqrt{0.55} Z_{2,t}$.

In Models~A-I, $Z_t=(Z_{1,t},Z_{2,t})^\tau$ is an independent centered stationary Gaussian process with covariance matrix $\Sigma=\left(\begin{matrix}
      1 & \rho\\ \rho & 1
    \end{matrix}\right)$. In each model, we consider two different choices of $\rho$, 0.1 and 0.5. For each of Models~C-E and G-I, we have standardized both components such that both of them have variance 1.

For each sample size $T=128,256,512,1024$ and each nominal size $\alpha=5\%$, 10\%, 15\%, Tables~\ref{tab:tab_comwithDKV} and \ref{tab:tab_comwithDKV2} present empirical rejection probabilities of test (\ref{eqn:testmethod_stationary}) for Models~A-E and F-I, respectively.

\begin{table}
    \caption{\label{tab:tab_comwithDKV} Rejection probabilities of tests (\ref{eqn:testmethod_stationary}) for the hypothesis of equal spectral density in models~A-E from simulated data.}
    \centering
 \begin{tabular*}{13.5cm}{@{\extracolsep{\fill}}rccccccc}
      \hline \noalign{\smallskip}
      $T$ & \multicolumn{3}{c}{$\rho=0.1$} & & \multicolumn{3}{c}{$\rho=0.5$}\\
      \cline{2-4} \cline{6-8}
      &  5\% & 10\% & 15\% &  & 5\% & 10\% & 15\% \\
      \noalign{\smallskip}\hline\noalign{\smallskip}
      & \multicolumn{7}{c}{Model A}\\
    128 & 0.044 & 0.081 & 0.133 & & 0.041 & 0.077 & 0.133\\
    256 & 0.067 & 0.131 & 0.181 & & 0.075 & 0.120 & 0.168\\
    512 & 0.067 & 0.120 & 0.171 & & 0.060 & 0.125 & 0.156\\
    1024 & 0.046 & 0.094 & 0.155 & & 0.039 & 0.090 & 0.140\\
     \noalign{\smallskip}
     & \multicolumn{7}{c}{Model B}\\
    128 & 0.056 & 0.104 & 0.150 & & 0.050 & 0.088 & 0.148\\
    256 & 0.056 & 0.111 & 0.168 & & 0.045 & 0.103 & 0.158\\
    512 & 0.066 & 0.106 & 0.155 & & 0.073 & 0.118 & 0.172\\
    1024 & 0.056 & 0.101 & 0.164 & & 0.050 & 0.097 & 0.145\\
     \noalign{\smallskip}
     & \multicolumn{7}{c}{Model C}\\ 
    128 & 0.050 & 0.098 & 0.140 & & 0.042 & 0.089 & 0.135\\
    256 & 0.065 & 0.127 & 0.176 & & 0.057 & 0.119 & 0.165\\
    512 & 0.061 & 0.110 & 0.162 & & 0.054 & 0.109 & 0.150\\
    1024 & 0.044 & 0.096 & 0.148 & & 0.044 & 0.098 & 0.137\\
      \noalign{\smallskip}
     & \multicolumn{7}{c}{Model D}\\ 
    128 & 0.046 & 0.095 & 0.130 & & 0.041 & 0.080 & 0.122\\
    256 & 0.058 & 0.121 & 0.172 & & 0.058 & 0.113 & 0.162\\
    512 & 0.060 & 0.121 & 0.160 & & 0.053 & 0.105 & 0.155\\
    1024 & 0.045 & 0.098 & 0.148 & & 0.037 & 0.091 & 0.137\\
      \noalign{\smallskip}
    & \multicolumn{7}{c}{Model E}\\ 
    128 & 0.056 & 0.118 & 0.193 & & 0.053 & 0.112 & 0.167\\
    256 & 0.056 & 0.093 & 0.148 & & 0.057 & 0.106 & 0.149\\
    512 & 0.030 & 0.077 & 0.121 & & 0.038 & 0.093 & 0.134\\
    1024 & 0.050 & 0.112 & 0.164 & & 0.064 & 0.111 & 0.166\\
\noalign{\smallskip}
 \hline
    \end{tabular*}
  \end{table}

Results in Table~\ref{tab:tab_comwithDKV} show that each empirical type I error rate is very close to the pre-specified nominal value. We conclude from Table~\ref{tab:tab_comwithDKV2} as follows. First, as $T$ increases, the performance of test (\ref{eqn:testmethod_stationary}) improves significantly. Second, the empirical power of test (\ref{eqn:testmethod_stationary}) has no clear relationship with the correlation $\rho$ of the process $Z_t$. The test statistic (\ref{eqn:proposedadstatistic}) has efficiently stripped the dependence between components of the time series. Third, the proposed test outperforms the competitive tests based on the Pearson statistic proposed by \cite{ZhangTu2018}. Models~F-I are also employed in the simulation studies of \cite{ZhangTu2018}, with identical settings. By comparing Table~\ref{tab:tab_comwithDKV2} with Tables~I and S.1 of \cite{ZhangTu2018}, we find for each parameter setting, the test (\ref{eqn:testmethod_stationary}) has higher empirical power than those in \cite{ZhangTu2018}. By our simulation experiments that are not reported here for space consideration, we also find that the empirical power of test (\ref{eqn:testmethod_stationary}) is quite robust when exchanging the order of components of the time series, i.e., replacing $(x_{1,t},x_{2,t})^\tau$ with $(x_{2,t},x_{1,t})^\tau$.
\begin{table}
    \caption{\label{tab:tab_comwithDKV2} Rejection probabilities of tests (\ref{eqn:testmethod_stationary}) for the hypothesis of equal spectral density in models~F-I from simulated data.}
    \centering
 \begin{tabular*}{13.5cm}{@{\extracolsep{\fill}}rccccccc}
      \hline \noalign{\smallskip}
      $T$ & \multicolumn{3}{c}{$\rho=0.1$} & & \multicolumn{3}{c}{$\rho=0.5$}\\
      \cline{2-4} \cline{6-8}
      &  5\% & 10\% & 15\% &  & 5\% & 10\% & 15\% \\
      \noalign{\smallskip}\hline\noalign{\smallskip}
    & \multicolumn{7}{c}{Model F}\\ 
    128 & 0.183 & 0.279 & 0.358 & & 0.197 & 0.279 & 0.347\\
    256 & 0.425 & 0.541 & 0.614 & & 0.417 & 0.533 & 0.606\\
    512 & 0.617 & 0.724 & 0.782 & & 0.614 & 0.727 & 0.783\\
    1024 & 0.835 & 0.899 & 0.929 & & 0.829 & 0.895 & 0.922\\
      \noalign{\smallskip}
     &  \multicolumn{7}{c}{Model G}\\
    128 & 0.128 & 0.228 & 0.299 & & 0.130 & 0.206 & 0.295\\
    256 & 0.285 & 0.410 & 0.488 & & 0.273 & 0.384 & 0.475\\
    512 & 0.443 & 0.566 & 0.628 & & 0.440 & 0.567 & 0.628\\
    1024 & 0.639 & 0.765 & 0.828 & & 0.642 & 0.750 & 0.809\\
     \noalign{\smallskip}
     & \multicolumn{7}{c}{Model H}\\ 
    128 & 0.089 & 0.162 & 0.224 & & 0.072 & 0.153 & 0.223\\
    256 & 0.117 & 0.209 & 0.286 & & 0.111 & 0.201 & 0.280\\
    512 & 0.138 & 0.235 & 0.321 & & 0.142 & 0.245 & 0.330\\
    1024 & 0.260 & 0.388 & 0.480 & & 0.256 & 0.396 & 0.504\\
      \noalign{\smallskip}
   & \multicolumn{7}{c}{Model I}\\ 
    128 & 0.129 & 0.200 & 0.254 & & 0.113 & 0.183 & 0.244\\
    256 & 0.166 & 0.242 & 0.304 & & 0.147 & 0.240 & 0.315\\
    512 & 0.182 & 0.281 & 0.354 & & 0.174 & 0.269 & 0.350\\
    1024 & 0.288 & 0.419 & 0.488 & & 0.320 & 0.428 & 0.507\\
      \noalign{\smallskip}
  \hline
    \end{tabular*}
  \end{table}

 To show the performance of test (\ref{eqn:testmethod_stationary}) for the models deviating from Gaussianity, we consider three stationary time series with Student $t$ innovations, i.e., Models~A$'$, B$'$ and F$'$. These three models admit the same expressions as Models~A, B and F, respectively. Although $Z_t=(Z_{1,t},Z_{2,t})^\tau$ is an independent centered process with covariance matrix $\Sigma=\left(\begin{matrix}
      1 & \rho\\ \rho & 1
    \end{matrix}\right)$, its components are non-Gaussian, in which $Z_{1,t}$ follows $t(d)$, a Student $t$ distribution with $d$ degrees of freedom, and $Z_{2,t}$ is defined by $Z_{2,t}=\rho Z_{1,t}+\sqrt{1-\rho^2} Z_{1,t}^\prime$, where $Z_{1,t}^\prime$ has the same distribution as $Z_{1,t}$ but is independent of $Z_{1,t}$. In each of these three models, we only consider the case of $\rho=0.5$, while varying the degrees of freedom from $d=5$ to 7.

For each sample size $T=128,256,512,1024$ and each nominal size $\alpha=5\%$, 10\%, 15\%, Presented in Table~\ref{tab:tab_performancefortinnovation} are empirical rejection probabilities of test (\ref{eqn:testmethod_stationary}) for Models~A$'$, B$'$ and F$'$. Table~\ref{tab:tab_performancefortinnovation} shows that each empirical type I error rate for Models~A$'$ and B$'$ is not as close as that for the counterparts, Models~A and B, but it is still very close to the pre-specified nominal one. Moreover, it becomes more close to the nominal one, with increasing the degree of freedom or increasing the sample size. From Tables~\ref{tab:tab_comwithDKV2} and \ref{tab:tab_performancefortinnovation}, it is also concluded that the empirical power of test (\ref{eqn:testmethod_stationary}) displays no clear difference when we keep the same expression of the model but change the innovation from Gaussian to non-Gaussian, or from $t(5)$ to $t(7)$.

 \begin{table}
    \caption{\label{tab:tab_performancefortinnovation} Rejection probabilities of tests (\ref{eqn:testmethod_stationary}) for the hypothesis of equal spectral density in models~A$'$, B$'$ and F$'$ from simulated data.}
    \centering
 \begin{tabular*}{13.5cm}{@{\extracolsep{\fill}}rccccccc}
      \hline \noalign{\smallskip}
      $T$ & \multicolumn{3}{c}{$d=5$} & & \multicolumn{3}{c}{$d=7$}\\
      \cline{2-4} \cline{6-8}
      &  5\% & 10\% & 15\% &  & 5\% & 10\% & 15\% \\
      \noalign{\smallskip}\hline\noalign{\smallskip}
      & \multicolumn{7}{c}{Model A$'$}\\
    128 & 0.077 & 0.174 & 0.209 & & 0.080 & 0.119 & 0.210\\
    256 & 0.100 & 0.161 & 0.215 & & 0.071 & 0.141 & 0.177\\
    512 & 0.098 & 0.154 & 0.207 & & 0.079 & 0.116 & 0.184\\
    1024 & 0.084 & 0.131 & 0.194 & & 0.065 & 0.126 & 0.170\\
     \noalign{\smallskip}
     & \multicolumn{7}{c}{Model B$'$}\\
    128 & 0.086 & 0.158 & 0.199 & & 0.062 & 0.134 & 0.171\\
    256 & 0.096 & 0.157 & 0.211 & & 0.072 & 0.104 & 0.181\\
    512 & 0.082 & 0.132 & 0.215 & & 0.060 & 0.105 & 0.165\\
    1024 & 0.080 & 0.149 & 0.197 & & 0.057 & 0.118 & 0.181\\
     \noalign{\smallskip}
    & \multicolumn{7}{c}{Model F$'$}\\ 
    128 & 0.217 & 0.314 & 0.386 & & 0.193 & 0.312 & 0.380\\
    256 & 0.425 & 0.537 & 0.609 & & 0.361 & 0.493 & 0.562\\
    512 & 0.612 & 0.731 & 0.785 & & 0.593 & 0.700 & 0.760\\
    1024 & 0.826 & 0.884 & 0.921 & & 0.846 & 0.908 & 0.934\\
\noalign{\smallskip}
 \hline
    \end{tabular*}
  \end{table}

\subsection{Comparing time-varying spectra}
\label{subsec:comp_timevaryingspec}

To demonstrate small-sample performance of test (\ref{eqn:testmethod_nonstationary}), we consider ten nonstationary time series $\mathbf{x}_{t, T}=(
  x_{1,t,T},x_{2,t,T}
)^\tau$, $t=1,\cdots,T$, in this section. Their components are as follows.

Model J (Copied smoothly-varying (SV) $MA(1)$ model.) $x_{i,t,T}=Z_{i,t}-\beta_1(t/T) Z_{i,t-1}$, $i=1,2$, where $\beta_1(u)=0.8(1+\sin(\pi u/2))$ for $u\in [0,1]$.

Model K (Copied $SV\,AR(1)$ model.) $x_{i,t,T}=\phi(t/T) x_{i,t-1,T} + Z_{i,t}$, $i=1,2$, where $\phi(u)=0.6\sin(4\pi u)$ for $u\in [0,1]$.

Model L (Copied wavelet process.) $x_{i,t,T}=\sum_{k=0}^{T-1} w_1(k/T) \psi_{1,k-t} Z_{i,k}$, $i=1,2$, where $w_1(u)=\cos(\pi u/2)$ for $u\in [0,1]$ and $\psi_{1,k}=\frac{1}{\sqrt{2}}(\mathbb{I}_{\{0\}}(k)-\mathbb{I}_{\{1\}}(k))$ for $k\in \mathbb{Z}$. This type of process was also considered by \cite{BellegemSachs2008}.

Model M (Copied Cholesky-decomposition model.) The model is given by
\begin{equation}
  \mathbf{x}_{t,T}=\sum_{k=1}^T \Phi(t/T,k/T)\exp(2\pi k t \mathrm{i}/T) \upepsilon_k, \label{eqn:GuoDai}
\end{equation}
where $\{\upepsilon_k, \, k=1,\cdots,n\}$ are independent. Moreover, for $k/T\neq 0, 0.5,1$, $\upepsilon_k$ follows a bivariate complex normal distribution with mean zero and covariance $n^{-1}\mathbf{I}_2$, and $\upepsilon_k=\overline{\upepsilon_{n-k}}$; for $k/T=0,0.5,1$, $\upepsilon_k$ follows a bivariate real normal distribution with mean zero and covariance $n^{-1}\mathbf{I}_2$. The process (\ref{eqn:GuoDai}) was constructed with the pre-specified spectral matrix $\mathbf{f}(u,\omega)=\Phi(u,\omega)\{\Phi(u,\omega)\}^\ast$ for $u\in [0,1]$ and $\omega\in (0,\pi)$ \citep[][Theorem~3.1]{GuoDai2006}. In model (\ref{eqn:GuoDai}), we set $\Phi(u,v)=(\psi_{ij}(u,v))_{2\times 2}=\mathrm{diag}(\psi_{11}(u,v),\psi_{22}(u,v))$ with $\psi_{11}(u,v)=\psi_{22}(u,v)=\{1.2 \cos(2\pi v)\}^2+0.3 \sin(2 \pi u)+0.7$.

Model N (Copied abruptly-varying $MA(2)$-$MA(2)$ model.) $x_{i,t,T}=Z_{i,t}-0.8 Z_{i,t-1}-(0.5-\gamma(t/T)) Z_{i,t-2}$, $i=1,2$, where $\gamma(u)=\mathbb{I}_{[0.6,1]}(u)$ for $u\in [0,1]$.

Model O ($SV\,MA(2)$-$SV\,MA(1)$ model.) $x_{1,t,T}=Z_{1,t}-\beta_1(t/T) Z_{1,t-1}-\beta_2(t/T) Z_{1,t-2}$ and $x_{2,t,T}=Z_{2,t}-\beta_1(t/T) Z_{2,t-1}$, where $\beta_1(u)$ is given as in Model~J and $\beta_2(u)=0.5 (1-\cos(\pi u))$ for $u\in [0,1]$.

Model P (Abruptly-varying $MA(2)$-$MA(2)$ model.) $x_{1,t,T}=Z_{1,t}-0.8 Z_{1,t-1}-(0.5-\gamma(t/T)) Z_{1,t-2}$ and $   x_{2,t,T}=Z_{2,t}-0.8 Z_{2,t-1}-0.5 Z_{1,t-2}$, where $\gamma(u)=\mathbb{I}_{[0.5,1]}(u)$ for $u\in [0,1]$.

Model Q ($SV\,AR(1)$-$SV\,AR(1)$ model.) $x_{1,t,T}=\phi(t/T) x_{1,t-1,T} + 1.5 Z_{1,t}$ and $x_{2,t,T}=\phi(t/T) x_{2,t-1,T} + Z_{2,t}$, where $\phi(u)$ is set as in Model~K.

Model R (Wavelet-wavelet process.) $x_{1,t,T}=\sum_{k=0}^{T-1} w_1(k/T) \psi_{1,k-t} Z_{1,k}$\\ $+\sum_{k=0}^{T-1} w_2(k/T) \psi_{2,k-t} Z_{2,k}$ and
$x_{2,t,T}=\sum_{k=0}^{T-1} w_1(k/T) \psi_{1,k-t} Z_{2,k}$, where $w_1(u)$ and $\psi_{1,k}$ are set as in Model~L, and $w_2(u)=0.3 u^2$ for $u\in [0,1]$ and $\psi_{2,k}=\frac{1}{2}(\mathbb{I}_{\{0,1\}}(k)-\mathbb{I}_{\{2,3\}}(k))$ for $k\in \mathbb{Z}$.

Model S (Cholesky-decomposition model.) The model is given by (\ref{eqn:GuoDai}) but we set $\psi_{11}(u,v)$ as in Model~M and $\psi_{22}(u,v)=\{1.2 \cos(2 \pi v)\}^2+0.6 \sin(2 \pi u)+0.7$. This choice makes $f_{22}(u,\omega)$ changes more rapidly over time than $f_{11}(u,\omega)$.

In Models~J-L and N-R, $Z_t=(Z_{1,t},Z_{2,t})^\tau$ and its covariance matrix are defined in the same way as in Models A-H. In each model, we only consider the case of $\rho=0.5$. For each model in Models~J-N, the two components have equal time-varying spectral densities, while for each model in Models~O-S, the two marginal spectral densities are unequal. For each model of Models~O and Q-S, both marginal spectra change smoothly over time; while for Model~N, both change abruptly, and Model~P, one changes smoothly and the other changes abruptly.

To illustrate that the test (\ref{eqn:testmethod_nonstationary}) still works well when applied to testing the null (\ref{eqn:nullhypothesis_equalityofspectraldensities}), we consider stationary Models~A, B and F, again. In each of these three models, we only consider the case of $\rho=0.5$.

For each sample size $T=128,256,512,1024$ and each nominal size $\alpha=5\%$, 10\% 15\%, Table~\ref{tab:tab_simulatedrejectionprobofcomptimevaryingspectra} presents empirical rejection probabilities of test (\ref{eqn:testmethod_nonstationary}) for Models~J-N and A-B, and Table~\ref{tab:tab_simulatedrejectionprobofcomptimevaryingspectra2} for Models~O-S and F.

\begin{table}
\caption{\label{tab:tab_simulatedrejectionprobofcomptimevaryingspectra} Rejection probabilities of test (\ref{eqn:testmethod_nonstationary}) for the hypothesis of equal time-varying spectral density in Models~J-N and A-B from simulated data.}
   \centering
   \begin{tabular*}{13.5cm}{@{\extracolsep{\fill}}rccccccccc}
      \hline \noalign{\smallskip}
      $T$ & $B$ & &  5\% & 10\% & 15\% &  & 5\% & 10\% & 15\% \\
      \noalign{\smallskip}\hline\noalign{\smallskip}
     & & & \multicolumn{3}{c}{Model~J} & & \multicolumn{3}{c}{Model~K}\\
      128 & 2 & & 0.057 & 0.119 & 0.176 & & 0.055 & 0.094 & 0.135\\
      256 & 3 & & 0.062 & 0.116 & 0.174 & & 0.055 & 0.106 & 0.158\\
      512 & 4 & & 0.066 & 0.132 & 0.180 & & 0.059 & 0.096 & 0.135\\
      1024 & 6 & & 0.054 & 0.109 & 0.161 & & 0.045 & 0.102 & 0.162\\
    \noalign{\smallskip}
   & & & \multicolumn{3}{c}{Model~L} & & \multicolumn{3}{c}{Model~M}\\
      128 & 2 & & 0.066 & 0.125 & 0.186 & & 0.048 & 0.100 & 0.166\\
      256 & 3 & & 0.106 & 0.163 & 0.223 & & 0.055 & 0.099 & 0.149\\
      512 & 4 & & 0.078 & 0.141 & 0.206 & & 0.055 & 0.102 & 0.153\\
      1024 & 6 & & 0.083 & 0.152 & 0.218 & & 0.038 & 0.093 & 0.154\\
    \noalign{\smallskip}
 & & & \multicolumn{3}{c}{Model~N} & & \multicolumn{3}{c}{Model~A}\\
      128 & 2 & & 0.053 & 0.140 & 0.167 & & 0.053 & 0.119 & 0.177\\
      256 & 3 & & 0.051 & 0.102 & 0.155 & & 0.062 & 0.126 & 0.190\\
      512 & 4 & & 0.059 & 0.120 & 0.167 & & 0.071 & 0.130 & 0.186\\
      1024 & 6 & & 0.045 & 0.114 & 0.151 & & 0.049 & 0.099 & 0.156\\
    \noalign{\smallskip}
    & & & \multicolumn{3}{c}{Model~B} & & \multicolumn{3}{c}{}\\
      128 & 2  & & 0.064 & 0.118 & 0.159\\
      256 & 3  & & 0.045 & 0.090 & 0.147\\
      512 & 4  & & 0.051 & 0.105 & 0.145\\
      1024 & 6  & & 0.057 & 0.123 & 0.171\\
    \noalign{\smallskip}
          \hline
    \end{tabular*}
  \end{table}

\begin{table}
\caption{\label{tab:tab_simulatedrejectionprobofcomptimevaryingspectra2} Rejection probabilities of test (\ref{eqn:testmethod_nonstationary}) for the hypothesis of equal time-varying spectral density in Models~O-S and F from simulated data.}
   \centering
   \begin{tabular*}{13.5cm}{@{\extracolsep{\fill}}rccccccccc}
      \hline \noalign{\smallskip}
      $T$ & $B$ & &  5\% & 10\% & 15\% &  & 5\% & 10\% & 15\% \\
      \noalign{\smallskip}\hline\noalign{\smallskip}
     & & & \multicolumn{3}{c}{Model~O} & & \multicolumn{3}{c}{Model~P}\\
      128 & 2 & & 0.114 & 0.204 & 0.277 & & 0.182 & 0.301 & 0.389\\
      256 & 3 & & 0.205 & 0.328 & 0.404 & & 0.168 & 0.288 & 0.366\\
      512 & 4 & & 0.353 & 0.492 & 0.590 & & 0.332 & 0.466 & 0.571\\
      1024 & 6 & & 0.514 & 0.662 & 0.753 & & 0.466 & 0.618 & 0.715\\
    \noalign{\smallskip}
   & & & \multicolumn{3}{c}{Model~Q} & & \multicolumn{3}{c}{Model~R}\\
      128 & 2 & & 0.529 & 0.658 & 0.732 & & 0.099 & 0.171 & 0.230\\
      256 & 3 & & 0.766 & 0.862 & 0.901 & & 0.249 & 0.360 & 0.443\\
      512 & 4 & & 0.955 & 0.988 & 0.991 & & 0.764 & 0.851 & 0.879\\
      1024 & 6 & & 0.999 & 1.000 & 1.000 & & 0.999 & 1.000 & 1.000\\
    \noalign{\smallskip}
   & & & \multicolumn{3}{c}{Model~S} & & \multicolumn{3}{c}{Model~F}\\
      128 & 2 & & 0.122 & 0.209 & 0.287 & & 0.117 & 0.185 & 0.261\\
      256 & 3 & & 0.163 & 0.265 & 0.344 & & 0.173 & 0.272 & 0.379\\
      512 & 4 & & 0.261 & 0.380 & 0.497 & & 0.307 & 0.442 & 0.533\\
      1024 & 6 & & 0.714 & 0.812 & 0.864 & & 0.408 & 0.586 & 0.692\\
    \noalign{\smallskip}
   \hline
    \end{tabular*}
  \end{table}

From Table~\ref{tab:tab_simulatedrejectionprobofcomptimevaryingspectra}, we can see that the empirical size is close to the nominal level for all models considered. From Table~\ref{tab:tab_simulatedrejectionprobofcomptimevaryingspectra2}, we can also observe that all deviations from equality of time-varying spectra are detected with reasonably large probabilities. Moreover, the proposed test outperforms the competitions based on the maximum of Pearson statistics proposed by \cite{ZhangTu2018}. Models~O-P and R-S were also employed in the simulation studies of \cite{ZhangTu2018}, with identical settings. By comparing Table~\ref{tab:tab_simulatedrejectionprobofcomptimevaryingspectra2} with Tables~II and S.2 of \cite{ZhangTu2018}, we find for each parameter setting, the test (\ref{eqn:testmethod_stationary}) has higher empirical power than its counterpart test proposed by \cite{ZhangTu2018}. When applied to testing the null (\ref{eqn:nullhypothesis_equalityofspectraldensities}), the statistic (\ref{eqn:teststatisticofnonstationarycase}) also outperforms the statistic (3.1) of \cite{ZhangTu2018}. It is unsurprising that the empirical power for Model~P is higher since the model does not satisfies the needed conditions in Theorem~\ref{thm:thm3.2nonstationary}. However, it is interesting to find that the empirical size is still close to the nominal level when the marginal spectral densities are equal but vary abruptly with respect to time $t$, by the simulation result of Model~N. Due to this, we make the conjecture that the result of Theorem~\ref{thm:thm3.2nonstationary} still holds even under some weaker conditions than the smoothing ones that given by (\ref{eqn:condnonstationary2}) and (\ref{eqn:condnonstationary3}).

\section{Conclusion}
\label{sec:conclusion}

In this paper, we proposed two test statistics, (\ref{eqn:proposedadstatistic}) and (\ref{eqn:teststatisticofnonstationarycase}), for comparing time-invariant and time-varying spectra. As in \cite{ZhangTu2018}, the test problems are transformed to the setting of goodness-of-fit tests. The test statistic (\ref{eqn:proposedadstatistic}) is constructed with the form as the A-D statistic. The test statistic (\ref{eqn:teststatisticofnonstationarycase}) is constructed by first computing  local A-D statistics and then maximizing them. Like the tests proposed by \cite{ZhangTu2018}, they have two advantages. First, it is easy to program and quite computationally efficient. Second, the proposed test statistic (\ref{eqn:teststatisticofnonstationarycase}) has a wide range of applicability in that it is applicable to stationary and locally stationary time series, with either independent or dependent components. Moreover, the proposed tests are independent of the partition of the sample space since they are based on the A-D statistic. In our simulation examples, we also find the proposed tests outperform those based on the Pearson statistic proposed by \cite{ZhangTu2018}.

Our recommended guidelines for implementation and extension are as follows:

1. Choice of $L$ and $B$. We recommend using the same scheme for the selection of tuning parameters $L$ and $B$, as those suggested in Section~6 of \cite{ZhangTu2018}.

2. Extension to weak dependence time series other than linear processes. By going through all our proofs, we find the main results in this paper would hold if the DFT $\mathbf{y}_T(\cdot)$ of the time series satisfies the following two conditions:
\begin{enumerate}
  \item[(i)] For any $K\in \mathbb{N}$ and $0<\omega_1<\cdots<\omega_K<\pi$, there exists a sequence of asymptotically independent random variables, $\{\upeta_T(\omega_k),\, k=1,\cdots,K\}$, satisfying that
     \begin{equation}
       \sup_{k=1,\cdots,K} \{\mathbf{E}|\mathbf{y}_T(\omega_k)-\upeta_T(\omega_k)|^2\}^{1/2} \leq C \frac{1}{\sqrt{T}},  \label{eqn:responseofclthm}
     \end{equation}
     where $C$ is independent $K$.
  \item[(ii)] For the class $\mathcal{A}$ of all measurable convex sets in $\mathbb{R}^4$, the inequality
      \begin{equation}
        \sup_{A\in \mathcal{A}}\Big|\mathbf{P}\Big(\big(\Re(\upeta_T(\omega_k)^\tau),\Im(\upeta_T(\omega_k)^\tau)\big)\in A \Big)-\Phi_k(A)\Big|\leq C\frac{1}{\sqrt{T}}, \label{eqn:responseofberry-esseen}
      \end{equation}
     holds uniformly for all $k=1,\cdots, K$, where $\Phi_k(A)$ is the probability of a $N\big(\mathbf{0},2^{-1} \mathbf{F}_k\big)$ random variable falling in $A$, with $\mathbf{F}_k=\mathrm{diag}\big(\mathbf{f}(\omega_k),\mathbf{f}(\omega_k)\big)$.
\end{enumerate}

For the linear process, the sequence of random variables satisfying conditions (i) and (ii), $\{\upeta_T(\omega_k),\, k=1,\cdots,K\}$, can be obtained by letting $\upeta_T(\omega)=\mathcal{A}(\omega) \mathbf{z}_T(\omega)$ for each $\omega\in \Omega_T:=\{2\pi k/T, k=0,1,\cdots,[T/2]\}$, where $\mathcal{A}(\omega)=(
  A_{ij}(\omega)
)_{2\times 2}=\sum_{n=-\infty}^\infty \Psi_n \mathrm{e}^{-\mathrm{i} n \omega}$ and $\mathbf{z}_T(\omega)=\frac{1}{\sqrt{2\pi T}}\sum_{t=1}^T Z_t \mathrm{e}^{\mathrm{i} \omega t}$. Due to the special expression of $\upeta_T(\omega)=\mathcal{A}(\omega) \mathbf{z}_T(\omega)$, the inequality (\ref{eqn:responseofberry-esseen}) is obtained by the Berry-Esseen theorem for independent random vectors~\citep[][eqn.~(1.5)]{Gotze1991}.

In the proof of our main results (Theorems~2.1 and 3.1), the expression of $\upeta_T(\omega)=\mathcal{A}(\omega) \mathbf{z}_T(\omega)$ plays a crucial role. However, the special expression relies on the linear representation of the time series $\{\mathbf{x}_t,\,t\in \mathbb{Z}\}$. It is an interesting problem to investigate whether there exists any nonlinear process, for which there exists a sequence of asymptotically independent random variables, $\{\upeta_T(\omega_k),\, k=1,\cdots,K\}$, satisfying both inequalities of (\ref{eqn:responseofclthm}) and  (\ref{eqn:responseofberry-esseen}). We think it would be reasonable to look for a dependence condition, e.g., that exerted on the physical dependence measure or the predictive dependence measure~\citep{Wu2005,Dahlhausetal2017}, under which there exists a sequence of asymptotically independent random variables, $\{\upeta_T(\omega_k),\, k=1,\cdots,K\}$, satisfying both inequalities of (\ref{eqn:responseofclthm}) and  (\ref{eqn:responseofberry-esseen}), but have not found such a typical condition yet. We think this is a challenging problem that need to be investigated further. Although the conditions, provided in Theorem~2.2 of \cite{Lahiri2003} can ensure that the DFTs are asymptotically independent, they can not guarantee that the properties (i) and (ii) hold. However, it can be concluded that under the conditions of Theorem~2.2 of \cite{Lahiri2003}, the asymptotic result (\ref{eqn:mainresultofthm1}) in Theorem~\ref{thm:thm3.1} still holds if first letting $T\rightarrow \infty$ and then letting $L\rightarrow \infty$.

3. Extension to multiple spectra. If $\{\mathbf{x}_t, t\in \mathbb{Z}\}$ is an $m$-dimensional ($m>2$) stationary process, the null hypothesis (\ref{eqn:nullhypothesis_equalityofspectraldensities}) is generalized to
\begin{equation*}
  H_0:\ f_{11}(\omega)=f_{22}(\omega)=\cdots=f_{mm}(\omega) \quad \textrm{on } (0,\pi).   \label{eqn:nullhypothesis_equalityofspectraldensitiesmulti}
\end{equation*}
Accordingly, $\hat{F}_{T,L} (x)$ defined in (\ref{eqn:hatFTL}) is replaced by
\begin{align*}
  \hat{F}_{T,L} (x)=&\frac{1}{L-1}\sum_{l=1}^{L-1} \Big\{\mathbb{I}_{(0,x]}\Big(\frac{I_{T,11}(\frac{(l-1)m+1-1/2}{L}\pi)}{I_{T,22}(\frac{(l-1)m+1}{L}\pi)}\Big)+\cdots\\
  &+\mathbb{I}_{(0,x]}\Big(\frac{I_{T,(m-1)(m-1)}(\frac{l m-1-1/2}{L}\pi)}{I_{T,mm}(\frac{l m-1}{L}\pi)}\Big)+\mathbb{I}_{(0,x]}\Big(\frac{I_{T,mm}(\frac{l m-1/2}{L}\pi)}{I_{T,11}(\frac{l m}{L}\pi)}\Big)\Big\} 
\end{align*}
for $x>0$, where $L\leq [T/(4m)]$. Then the test statistic (\ref{eqn:proposedadstatistic}) can be generalized to the case with $m>2$. By applying similar extensions to $\hat{F}_{M,L}^{(k)} (x)$, the test statistic (\ref{eqn:teststatisticofnonstationarycase}) is also readily extended to more than two spectra.

\section*{Acknowledgements}

This work was supported by the National Natural Science Foundation of China (grant number: 11671416, 11971116) and the Natural Science Foundation of Shanghai (grant number: 20JC1413800).



\appendix

\renewcommand{\theequation}{A.\arabic{equation}}
\setcounter{equation}{0}

\section*{Appendix. Proofs}
\label{sec:appendix}

In this appendix, we prove the results from Section~\ref{sec:comparingstationaryspectra}.

\subsection*{A.1.  Proof of Theorem~\ref{thm:thm3.1}}

For notational brevity, we write $\ell_l$ and $\ell_l^\prime$ for $\frac{l-1/2}{L}\pi$ and $\frac{l}{L}\pi$, respectively.

Let $\upeta_T(\omega)=(\eta_T^{(1)}(\omega),\eta_T^{(2)}(\omega))^\tau=\mathcal{A}(\omega) \mathbf{z}_T(\omega)$ for $\omega=2\pi k/T$, $k=0,1,\cdots,[T/2]$, where
$\mathcal{A}(\omega)$ is defined in (\ref{eqn:fomegadetail}), and $\mathbf{z}_T(\omega)=\frac{1}{\sqrt{2\pi T}}\sum_{t=1}^T Z_t \mathrm{e}^{\mathrm{i} \omega t}$. Then we extend the definition of $\upeta_T(\cdot)$ to a piecewise constant function on $[-\pi,\pi]$ as in (\ref{eqn:extentionmathbfy}). For each integer $L>1$, we define
\begin{equation*}
  \tilde{A}_{T,L}=(L-1) \int_0^\infty \frac{(\tilde{F}_{T,L}(x)-F(x))^2}{F(x)(1-F(x))}\,\mathrm{d}F(x)
\end{equation*}
and
\begin{equation*}  \breve{A}_{T,L}=(L-1) \int_0^\infty \frac{(\breve{F}_{T,L}(x)-F(x))^2}{F(x)(1-F(x))}\,\mathrm{d}F(x),
\end{equation*}
where
\begin{equation*}
  \tilde{F}_{T,L}(x)=\frac{1}{L-1}\sum_{l=1}^{L-1} \mathbb{I}_{(0,x]}\Big(\frac{I_{T,11}(\ell_l)/f_{11}(\ell_l)}{I_{T,22}(\ell_l^\prime)/f_{22}(\ell_l^\prime)}\Big)
\end{equation*}
and
\begin{equation*}
\breve{F}_{T,L}(x)=\frac{1}{L-1}\sum_{l=1}^{L-1} \mathbb{I}_{(0,x]}\Big(\frac{|\eta_T^{(1)}(\ell_l)|^2/f_{11}(\ell_l)}{|\eta_T^{(2)}(\ell_l^\prime)|^2/f_{22}(\ell_l^\prime)}\Big)  \end{equation*}
for $x\in(0,\infty)$, respectively.

We prove in Appendix~A.2 that
\begin{equation}
  \tilde{A}_{T,L}=\breve{A}_{T,L}+ o_p\big(\frac{\sqrt{L \log T}}{\sqrt{T}}\big) \label{eqn:arguentofthm3.1-4}
\end{equation}
and
\begin{equation}
  \hat{A}_{T,L}=\tilde{A}_{T,L}+ o_p\big(\frac{\sqrt{L \log T}}{\sqrt{T}}\big) \label{eqn:arguentofthm3.1-5}
\end{equation}
hold as $T$ goes to infinity, where $o_p(T)$ means $o_p(T)/T$ converges to zero in probability.

Combining (\ref{eqn:arguentofthm3.1-4}) with (\ref{eqn:arguentofthm3.1-5}), we obtain
\begin{equation}
\hat{A}_{T,L}=\breve{A}_{T,L}+ o_p\big(\frac{\sqrt{L \log T}}{\sqrt{T}}\big).\label{eqn:arguentofthm3.1-205}
\end{equation}

Let $\phi_{1,T}(\omega)=|\eta_T^{(1)}(\omega)|^2/f_{11}(\omega)$ and let $\phi_{2,T}(\omega)=|\eta_T^{(2)}(\omega)|^2/f_{22}(\omega)$. According to the arguments in the proof of Lemma~2.1 of \cite{ZhangTu2018}, for each $l=1,\cdots,L-1$, $\big(\phi_{1,T}(\ell_l),\phi_{2,T}(\ell_l^\prime)\big)^\tau$ is a function of $\big(\eta_T^{(1)}(\ell_l),\eta_T^{(2)}(\ell_l^\prime)\big)^\tau$ that is an asymptotically zero-mean complex normal random variable with covariance matrix $\mathrm{diag}(f_{11}(\ell_l), f_{22}(\ell_l^\prime))$. Moreover, $\big(|\eta_T^{(1)}(\ell_l)|^2/f_{11}(\ell_l)$, $|\eta_T^{(2)}(\ell_l^\prime)|^2/f_{22}(\ell_l^\prime)\big)^\tau$ is asymptotic independent among $l=1,\cdots,L-1$. Then we have that $\{\phi_{1,T}(\ell_l), \, l=1,\cdots,L-1\}$ and $\{\phi_{2,T}(\ell_l^\prime), \, l=1,\cdots,L-1\}$ are two sequences of asymptotic $\chi^2(2)$-distributed random variables. In view of Skorohod's theorem~\citep[e.g.][Theorem~25.6]{Billingsley1995}, for each $L$, there exist random variables $\{\phi_{1,T,l},\,l=1,\cdots,L-1\}$, $\{\phi_{2,T,l},\,l=1,\cdots,L-1\}$, $\{\chi_{1,l}^2,\,l=1,\cdots,L-1\}$ and $\{\chi_{2,l}^2,\,l=1,\cdots,L-1\}$ defined on a common probability space such that (a) $\phi_{1,T}(\ell_l)\stackrel{d}{=}\phi_{1,T,l}$ and $\phi_{2,T}(\ell_l^\prime)\stackrel{d}{=}\phi_{2,T,l}$ for each $T$ and $l$, and (b) $\phi_{1,T,l}\stackrel{a.s.}{\longrightarrow} \chi_{1,l}^2$ and $\phi_{2,T,l}\stackrel{a.s.}{\longrightarrow} \chi_{2,l}^2$ for each $l$, as $T$ goes to infinity, where $\chi_{1,l}^2$, $\chi_{2,l}^2$, $l=1,\cdots,L-1$, are independent $\chi^2(2)$-distributed random variables, ``$\stackrel{d}{=}$'' denotes equality in distribution, and ``$\stackrel{a.s.}{\longrightarrow}$'' denotes convergence almost surely. For each integer $L>1$, we denote
\begin{equation*}
  \breve{A}^\ast_{T,L}=(L-1) \int_0^\infty \frac{(\breve{F}^\ast_{T,L}(x)-F(x))^2}{F(x)(1-F(x))}\,\mathrm{d}F(x)
\end{equation*}
and
\begin{equation}  \breve{A}^\ast_{L}=(L-1) \int_0^\infty \frac{(\breve{F}^\ast_{L}(x)-F(x))^2}{F(x)(1-F(x))}\,\mathrm{d}F(x),
\label{eqn:breveAastL}
\end{equation}
where $\breve{F}^\ast_{T,L}(x)=\frac{1}{L-1}\sum_{l=1}^{L-1} \mathbb{I}_{(0,x]}\big(\phi_{1,T,l}/\phi_{2,T,l}\big)$ and $\breve{F}^\ast_{L}(x)=\frac{1}{L-1}\sum_{l=1}^{L-1} \mathbb{I}_{(0,x]}\big(\chi_{1,l}^2/\chi_{2,l}^2\big)$
for $x\in(0,\infty)$, respectively. We prove in Appendix~A.2 that
\begin{equation}
\breve{A}^\ast_{T,L}=\breve{A}^\ast_{L}+o_p(1)
\label{eqn:mainresultofthm1-3inproof}
\end{equation}
holds uniformly for $L>1$, as $T$ goes to infinity.
 Now it holds that \begin{equation}
  \breve{A}_{T,L}\stackrel{d}{=} \breve{A}^\ast_{T,L} \quad \textrm{and}\quad \breve{A}^\ast_{T,L}=\breve{A}^\ast_{L}+o_p(1)
  \label{eqn:mainresultofthm1-2inproof}
\end{equation}
uniformly for $L>1$.

Since $\breve{A}^\ast_{L}$ is defined by (\ref{eqn:breveAastL}), according to Theorem~26.1 of \cite{DasGupta2008}, we have
\begin{equation}
    \breve{A}^\ast_{L} \rightsquigarrow \int_0^1 \frac{B_0^2(t)}{t(1-t)}\,\mathrm{d}t, \label{eqn:mainresultofthm1inproof}
  \end{equation}
as $L$ goes to infinity.

Combining (\ref{eqn:arguentofthm3.1-205}), (\ref{eqn:mainresultofthm1-2inproof}) and (\ref{eqn:mainresultofthm1inproof}) produces (\ref{eqn:mainresultofthm1}). This completes the proof of Theorem~\ref{thm:thm3.1}. 

\subsection*{A.2. Some technical details}

This appendix contains details in verifying (\ref{eqn:arguentofthm3.1-4}),  (\ref{eqn:arguentofthm3.1-5}) and (\ref{eqn:mainresultofthm1-3inproof}).

Note that
\begin{align}
  \tilde{A}_{T,L}=&(L-1) \int_0^\infty\frac{(\tilde{F}_{T,L}(x)-\breve{F}_{T,L}(x))^2}{F(x)(1-F(x))}\,\mathrm{d}F(x) \nonumber\\
  &+2\int_0^\infty \frac{(\sqrt{L-1}(\tilde{F}_{T,L}(x)-\breve{F}_{T,L}(x)))(\sqrt{L-1}(\breve{F}_{T,L}(x)-F(x)))}{F(x)(1-F(x))}\,\mathrm{d}F(x) \nonumber\\
  &+\breve{A}_{T,L}. \label{eqn:proofofthm-12}
\end{align}

To prove (\ref{eqn:arguentofthm3.1-4}), it suffices to verify
\begin{equation}
   \sup_{x>0}\sqrt{L-1}\big|\tilde{F}_{T,L}(x)-\breve{F}_{T,L}(x)\big|=o_p\big(\frac{\sqrt{L \log T}}{\sqrt{T}}\big) \label{eqn:arguentofthm2.1-4}
\end{equation}
and
\begin{equation}
  \int_0^\infty\frac{\sqrt{L-1}(\tilde{F}_{T,L}(x)-\breve{F}_{T,L}(x))}{F(x)(1-F(x))}\,\mathrm{d}F(x)
  =o_p\big(\frac{\sqrt{L \log T}}{\sqrt{T}}\big). \label{eqn:arguentofthm2.1-8}
\end{equation}

If (\ref{eqn:arguentofthm2.1-4}) and (\ref{eqn:arguentofthm2.1-8}) hold, we have
\begin{align}
  &(L-1)\int_0^\infty\frac{(\tilde{F}_{T,L}(x)-\breve{F}_{T,L}(x))^2}{F(x)(1-F(x))}\,\mathrm{d}F(x) =\int_0^\infty\frac{(\sqrt{L-1}(\tilde{F}_{T,L}(x)-\breve{F}_{T,L}(x)))^2}{F(x)(1-F(x))}\,\mathrm{d}F(x)
  \nonumber  \\
  &\qquad \leq o_p\big(\frac{\sqrt{L \log T}}{\sqrt{T}}\big)\int_0^\infty \frac{\sqrt{L-1}(\tilde{F}_{T,L}(x)-\breve{F}_{T,L}(x))}{F(x)(1-F(x))}\,\mathrm{d}F(x)=o_p\big(\frac{L \log T}{T}\big).\label{eqn:proofofthm-11} \end{align}
Moreover, by the Cauchy inequality, we have
\begin{align}
  &\Big|\int_0^\infty\frac{(\sqrt{L-1}(\tilde{F}_{T,L}(x)-\breve{F}_{T,L}(x)))(\sqrt{L-1}(\breve{F}_{T,L}(x)-F(x)))}{F(x)(1-F(x))}\,\mathrm{d}F(x)\Big|
  \nonumber\\
  &\leq  \sqrt{(L-1)\int_0^\infty\frac{(\tilde{F}_{T,L}(x)-\breve{F}_{T,L}(x))^2}{F(x)(1-F(x))}\,\mathrm{d}F(x) \,\breve{A}_{T,L}}=o_p\big(\frac{\sqrt{L \log T}}{\sqrt{T}}\big).\label{eqn:proofofthm-13}
\end{align}
Combining (\ref{eqn:proofofthm-12}), (\ref{eqn:proofofthm-11}) and (\ref{eqn:proofofthm-13}), we obtain (\ref{eqn:arguentofthm3.1-4}).

Similarly, to prove (\ref{eqn:arguentofthm3.1-5}), it suffices to verify
\begin{equation}
   \sup_{x>0}\sqrt{L-1}|\hat{F}_{T,L}(x)-\tilde{F}_{T,L}(x)|=o_p\big(\frac{\sqrt{L \log T}}{\sqrt{T}}\big) \label{eqn:arguentofthm2.1-104}
\end{equation}
and
\begin{equation}
  \int_0^\infty\frac{\sqrt{L-1}(\hat{F}_{T,L}(x)-\tilde{F}_{T,L}(x))}{F(x)(1-F(x))}\,\mathrm{d}F(x)
  =o_p\big(\frac{\sqrt{L \log T}}{\sqrt{T}}\big); \label{eqn:arguentofthm2.1-108}
\end{equation}
and to prove (\ref{eqn:mainresultofthm1-3inproof}), it suffices to check
\begin{equation}
   \sup_{x>0}\sqrt{L-1}|\breve{F}^\ast_{T,L}(x)-\breve{F}^\ast_{L}(x)|=o_p\big(1\big) \label{eqn:arguentofthm2.1-504}
\end{equation}
and
\begin{equation}
  \int_0^\infty\frac{\sqrt{L-1}(\breve{F}^\ast_{T,L}(x)-\breve{F}^\ast_{L}(x))}{F(x)(1-F(x))}\,\mathrm{d}F(x)
  =o_p\big(1\big). \label{eqn:arguentofthm2.1-508}
\end{equation}

The rest of this appendix pertains to verification of (\ref{eqn:arguentofthm2.1-4}), (\ref{eqn:arguentofthm2.1-8}), (\ref{eqn:arguentofthm2.1-104}), (\ref{eqn:arguentofthm2.1-108}), (\ref{eqn:arguentofthm2.1-504}) and (\ref{eqn:arguentofthm2.1-508}).

\textbf{Proof of (\ref{eqn:arguentofthm2.1-4}) and (\ref{eqn:arguentofthm2.1-104}).} It suffices to check
\begin{equation}
 \max_{l=1,\cdots,L-1} \mathbf{E}\Big|\mathbb{I}_{(0,x]}\Big(\frac{I_{T,11}(\ell_l)/f_{11}(\ell_l)}{I_{T,22}(\ell_l^\prime)/f_{22}(\ell_l^\prime)}\Big)-\mathbb{I}_{(0,x]}\Big(\frac{|\eta_T^{(1)}(\ell_l)|^2/f_{11}(\ell_l)}{|\eta_T^{(2)}(\ell_l^\prime)|^2/f_{22}(\ell_l^\prime)}\Big)\Big|\leq C\frac{1}{\sqrt{T}} \label{eqn:arguentofthm2.1-20}
\end{equation}
and
\begin{equation}
 \max_{l=1,\cdots,L-1} \mathbf{E}\Big|\mathbb{I}_{(0,x]}\Big(\frac{I_{T,11}(\ell_l)/f_{11}(\ell_l)}{I_{T,22}(\ell_l^\prime)/f_{22}(\ell_l^\prime)}\Big)-\mathbb{I}_{(0,x]}\Big(\frac{I_{T,11}(\ell_l)}{I_{T,22}(\ell_l^\prime)}\Big)\Big|\leq C\frac{1}{\sqrt{T}}. \label{eqn:arguentofthm2.2-3}
\end{equation}
hold for all $x>0$.

For instance, if the inequality (\ref{eqn:arguentofthm2.1-20}) holds, by the Markov inequality and the triangular inequality, we have for all $\epsilon>0$,
\begin{align*}
  &\mathbf{P}\Big(\frac{\big|\sqrt{L-1}\tilde{F}_{T,L}(x)-\sqrt{L-1}\breve{F}_{T,L}(x)\big|}{\sqrt{L\log T}/\sqrt{T}}>\epsilon \Big)\\
  &\quad \leq \frac{\sqrt{T}}{\epsilon\sqrt{L}\log T} \frac{1}{\sqrt{L-1}} \sum_{l=1}^{L-1}
  \mathbf{E}\Big|\mathbb{I}_{(0,x]}\Big(\frac{I_{T,11}(\ell_l)/f_{11}(\ell_l)}{I_{T,22}(\ell_l^\prime)/f_{22}(\ell_l^\prime)}\Big)-\mathbb{I}_{(0,x]}\Big(\frac{|\eta_T^{(1)}(\ell_l)|^2/f_{11}(\ell_l)}{|\eta_T^{(2)}(\ell_l^\prime)|^2/f_{22}(\ell_l^\prime)}\Big)\Big|\\
  &\quad \leq  C \frac{\sqrt{T}}{\epsilon\sqrt{L\log T}} \sqrt{L-1} \frac{1}{\sqrt{T}}\longrightarrow 0
\end{align*}
as $T$ goes to infinity. This proves (\ref{eqn:arguentofthm2.1-4}).

Note that both inequalities of (\ref{eqn:arguentofthm2.1-20}) and (\ref{eqn:arguentofthm2.2-3}) hold, since they are the special cases of inequalities (S.9) and (S.10) in the supporting information of \cite{ZhangTu2018} when $a=0$ and $b=x$, respectively. This proves (\ref{eqn:arguentofthm2.1-20}) and (\ref{eqn:arguentofthm2.2-3}).

\textbf{Proof of (\ref{eqn:arguentofthm2.1-8}).} According to (S.5)-(S.8) in the supporting information of \cite{ZhangTu2018}, we have the decompositions
\begin{equation*}
  I_{T,11}(\ell_l)=|\eta_T^{(1)}(\ell_l)|^2+\vartheta_T^{(1)}(\ell_l) \quad \textrm{and}\quad I_{T,22}(\ell_l^\prime)=|\eta_T^{(2)}(\ell_l^\prime)|^2+\vartheta_T^{(2)}(\ell_l^\prime),
\end{equation*}
where
\begin{equation}
 \vartheta_T^{(1)}(\ell_l)=o_p \big(\frac{\sqrt{\log T}}{\sqrt{T}}\big) \quad \textrm{ and } \quad \vartheta_T^{(2)}(\ell_l^\prime)=o_p \big(\frac{\sqrt{\log T}}{\sqrt{T}}\big) \label{eqn:arguentofthm2.1-50}
\end{equation}
hold for all $l=1,\cdots,L$. Note that
\begin{align}
 &\int_0^\infty\frac{\sqrt{L-1}(\tilde{F}_{T,L}(x)-\breve{F}_{T,L}(x))}{F(x)(1-F(x))}\,\mathrm{d}F(x)
 = \sqrt{L-1}\int_0^\infty (\tilde{F}_{T,L}(x)-\breve{F}_{T,L}(x)) x^{-1} \,\mathrm{d} x \nonumber\\
 &\qquad =\frac{1}{\sqrt{L-1}}\sum_{l=1}^{L-1} \int_0^\infty \frac{\mathbb{I}_{(0,x]}\big(\frac{I_{T,11}(\ell_l)/f_{11}(\ell_l)}{I_{T,22}(\ell_l^\prime)/f_{22}(\ell_l^\prime)}\big)-\mathbb{I}_{(0,x]}\big(\frac{|\eta_T^{(1)}(\ell_l)|^2/f_{11}(\ell_l)}{|\eta_T^{(2)}(\ell_l^\prime)|^2/f_{22}(\ell_l^\prime)}\big)}{x} \,\mathrm{d} x. \label{eqn:arguentofthm2.1-51}
\end{align}
Since $|\log(1+x)|=O(x)$ as $x$ goes to zero, it follows from (\ref{eqn:arguentofthm2.1-50}) that
\begin{align}
 & \Big|\int_0^\infty \frac{\mathbb{I}_{(0,x]}\big(\frac{I_{T,11}(\ell_l)/f_{11}(\ell_l)}{I_{T,22}(\ell_l^\prime)/f_{22}(\ell_l^\prime)}\big)-\mathbb{I}_{(0,x]}\big(\frac{|\eta_T^{(1)}(\ell_l)|^2/f_{11}(\ell_l)}{|\eta_T^{(2)}(\ell_l^\prime)|^2/f_{22}(\ell_l^\prime)}\big)}{x} \,\mathrm{d} x\Big| \nonumber\\
 &\qquad=\Big|\log \Big(\frac{I_{T,11}(\ell_l)/f_{11}(\ell_l)}{I_{T,22}(\ell_l^\prime)/f_{22}(\ell_l^\prime)}\Big)
 -\log \Big(\frac{|\eta_T^{(1)}(\ell_l)|^2/f_{11}(\ell_l)}{|\eta_T^{(2)}(\ell_l^\prime)|^2/f_{22}(\ell_l^\prime)}\Big)
 \Big| \nonumber\\
 &\qquad=\Big|\log \Big(\frac{I_{T,11}(\ell_l)/f_{11}(\ell_l)}{|\eta_T^{(1)}(\ell_l)|^2/f_{11}(\ell_l)}\Big)
 -\log \Big(\frac{I_{T,22}(\ell_l^\prime)/f_{22}(\ell_l^\prime)}{|\eta_T^{(2)}(\ell_l^\prime)|^2/f_{22}(\ell_l^\prime)}\Big)
 \Big| \nonumber\\
 &\qquad =\Big|\log\Big(1+\frac{\vartheta_T^{(1)}(\ell_l)/f_{11}(\ell_l)}{|\eta_T^{(1)}(\ell_l)|^2/f_{11}(\ell_l)}\Big) -\log\Big(1+\frac{\vartheta_T^{(2)}(\ell_l^\prime)/f_{22}(\ell_l^\prime)}{|\eta_T^{(2)}(\ell_l^\prime)|^2/f_{22}(\ell_l^\prime)}\Big) \Big| \nonumber\\
 &\qquad \leq \Big|\log\Big(1+\frac{\vartheta_T^{(1)}(\ell_l)/f_{11}(\ell_l)}{|\eta_T^{(1)}(\ell_l)|^2/f_{11}(\ell_l)}\Big) \Big|+\Big|\log\Big(1+\frac{\vartheta_T^{(2)}(\ell_l^\prime)/f_{22}(\ell_l^\prime)}{|\eta_T^{(2)}(\ell_l^\prime)|^2/f_{22}(\ell_l^\prime)}\Big) \Big|\nonumber\\
 &\qquad =o_p\big(\frac{\sqrt{\log T}}{\sqrt{T}}\big) \label{eqn:arguentofthm2.1-52}
\end{align}
holds for all $x>0$ and $l=1,\cdots,L$. Combining (\ref{eqn:arguentofthm2.1-51}) and (\ref{eqn:arguentofthm2.1-52}), we obtain (\ref{eqn:arguentofthm2.1-8}).

\textbf{Proof of (\ref{eqn:arguentofthm2.1-108}).} By the Lipschitz continuity, we have $|f_{22}(\ell_l)-f_{22}(\ell_l^\prime)|\leq C \frac{1}{L}\leq C\frac{1}{\sqrt{T}}=o(\frac{\log T}{\sqrt{T}})$.
Then, if the null hypothesis (\ref{eqn:nullhypothesis_equalityofspectraldensities}) is true, we obtain \begin{align*}
 & \Big|\log \Big(\frac{I_{T,11}(\ell_l)/f_{11}(\ell_l)}{I_{T,22}(\ell_l^\prime)/f_{22}(\ell_l^\prime)}\Big)
 -\log \Big(\frac{I_{T,11}(\ell_l)}{I_{T,22}(\ell_l^\prime)}\Big)
 \Big|=\Big|\log \Big(\frac{f_{22}(\ell_l^\prime)}{f_{11}(\ell_l)}\Big)
\Big| \nonumber\\
 &\qquad =\Big|\log\Big(1+\frac{f_{22}(\ell_l^\prime)-f_{22}(\ell_l)}{f_{22}(\ell_l)}\Big)\Big|=o\big(\frac{\sqrt{\log T}}{\sqrt{T}}\big)
\end{align*}
holds for all $x>0$ and $l=1,\cdots,L$, since $f_{22}(\omega)$ is lower bounded. With arguments similar to the proof of (\ref{eqn:arguentofthm2.1-8}), we obtain (\ref{eqn:arguentofthm2.1-108}).

\textbf{Proof of (\ref{eqn:arguentofthm2.1-504}).} To prove (\ref{eqn:arguentofthm2.1-504}), by the Chebyshev inequality, it suffices to verify
\begin{equation}
   \sup_{x>0}\mathbf{E}[(L-1)(\breve{F}^\ast_{T,L}(x)-\breve{F}^\ast_{L}(x))^2]=o\big(1\big). \label{eqn:arguentofthm2.1-509}
\end{equation}

Note that \begin{align}
 & \mathbf{E}\big[(L-1)(\breve{F}^\ast_{T,L}(x)-\breve{F}^\ast_{L}(x))^2\big]
  =\frac{1}{L-1}\sum_{l=1}^{L-1} \mathbf{E}\Big[\Big( \mathbb{I}_{(0,x]}\Big(\frac{\phi_{1,T,l}}{\phi_{2,T,l}}\Big) -\mathbb{I}_{(0,x]}\Big(\frac{\chi^2_{1,l}}{\chi^2_{2,l}}\Big)\Big)^2\Big]
  \nonumber\\
  &+\frac{2}{L-1}\sum_{l_1=1}^{L-2} \sum_{l_2=l_1+1}^{L-1} \mathbf{E}\Big[\Big(\mathbb{I}_{(0,x]}\Big(\frac{\phi_{1,T,l_1}}{\phi_{2,T,l_1}}\Big)
  -\mathbb{I}_{(0,x]}\Big(\frac{\chi^2_{1, l_1}}{\chi^2_{2, l_1}}\Big)\Big)
  \Big(\mathbb{I}_{(0,x]}\Big(\frac{\phi_{1,T,l_2}}{\phi_{2,T,l_2}}\Big)-\mathbb{I}_{(0,x]}\Big(\frac{\chi^2_{1, l_2}}{\chi^2_{2, l_2}}\Big)\Big)\Big].
  \label{eqn:arguentofthm2.1-R1-14}
\end{align}
By the perfect square formula, we have
\begin{align}
  &\mathbf{E}\Big[\Big( \mathbb{I}_{(0,x]}\Big(\frac{\phi_{1,T,l}}{\phi_{2,T,l}}\Big) -\mathbb{I}_{(0,x]}\Big(\frac{\chi^2_{1,l}}{\chi^2_{2,l}}\Big)\Big)^2\Big]\nonumber\\
 & = \mathbf{P}\Big(\frac{\phi_{1,T,l}}{\phi_{2,T,l}}\leq x\Big)
   +\mathbf{P}\Big(\frac{\chi^2_{1,l}}{\chi^2_{2,l}}\leq x\Big)
   -2\, \mathbf{P}\Big(\frac{\phi_{1,T,l}}{\phi_{2,T,l}}\leq x, \frac{\chi^2_{1,l}}{\chi^2_{2,l}}\leq x\Big).
   \label{eqn:arguentofthm2.1-R1-15}
\end{align}
By the Berry-Esseen theorem for random vectors~\cite[][eqn.~(1.5)]{Gotze1991}, we obtain that
\begin{equation}
 \mathbf{P}\Big(\frac{\phi_{1,T,l}}{\phi_{2,T,l}}\leq x\Big)=\mathbf{P}\Big(\frac{\chi^2_{1,l}}{\chi^2_{2,l}}\leq x\Big)+o\big(\frac{\sqrt{\log T}}{\sqrt{T}}\big)
\label{eqn:arguentofthm2.1-R1-16}
\end{equation}
and
\begin{equation}\mathbf{P}\Big(\frac{\phi_{1,T,l}}{\phi_{2,T,l}}\leq x, \frac{\chi^2_{1,l}}{\chi^2_{2,l}}\leq x\Big)=\mathbf{P}\Big(\frac{\chi^2_{1,l}}{\chi^2_{2,l}}\leq x\Big)+o\big(\frac{\sqrt{\log T}}{\sqrt{T}}\big)
\label{eqn:arguentofthm2.1-R1-17}
\end{equation}
hold uniformly for $x>0$~\citep[cf.][eqn.~(2.14)]{ZhangTu2018}, where the equality (\ref{eqn:arguentofthm2.1-R1-17}) is derived by employing the theorem of total probability according as the sign of $\phi_{i,T,l}-\chi^2_{i,l}$. It follows from Exercise~2.13.22, Theorems~4.3.1 and 4.4.1 of \cite{Brillinger2001} that $o\big(\frac{\sqrt{\log T}}{\sqrt{T}}\big)$ in (\ref{eqn:arguentofthm2.1-R1-16}) and (\ref{eqn:arguentofthm2.1-R1-17}) can be taken uniformly for $l=1,\cdots,L-1$. Combining (\ref{eqn:arguentofthm2.1-R1-15}), (\ref{eqn:arguentofthm2.1-R1-16}) and (\ref{eqn:arguentofthm2.1-R1-17}) gives
\begin{equation}
   \mathbf{E}\Big[\Big( \mathbb{I}_{(0,x]}\Big(\frac{\phi_{1,T,l}}{\phi_{2,T,l}}\Big) -\mathbb{I}_{(0,x]}\Big(\frac{\chi^2_{1,l}}{\chi^2_{2,l}}\Big)\Big)^2\Big]
   =o\big(\frac{\sqrt{\log T}}{\sqrt{T}}\big).
   \label{eqn:arguentofthm2.1-R1-18}
\end{equation}
Similarly, we obtain that
\begin{equation}
 \mathbf{E}\Big[\Big(\mathbb{I}_{(0,x]}\Big(\frac{\phi_{1,T,l_1}}{\phi_{2,T,l_1}}\Big)
  -\mathbb{I}_{(0,x]}\Big(\frac{\chi^2_{1, l_1}}{\chi^2_{2, l_1}}\Big)\Big)
  \Big(\mathbb{I}_{(0,x]}\Big(\frac{\phi_{1,T,l_2}}{\phi_{2,T,l_2}}\Big)-\mathbb{I}_{(0,x]}\Big(\frac{\chi^2_{1, l_2}}{\chi^2_{2, l_2}}\Big)\Big)\Big]=o\big(\frac{\log T}{T}\big)
  \label{eqn:arguentofthm2.1-R1-19}
\end{equation}
holds uniformly for $x>0$ and $l=1,\cdots,L-1$~\cite[cf.][the proof of Theorem~5.2.4]{Brillinger2001}. Combining (\ref{eqn:arguentofthm2.1-R1-14}), (\ref{eqn:arguentofthm2.1-R1-18}) and (\ref{eqn:arguentofthm2.1-R1-19}), we obtain
\begin{equation*}
   \mathbf{E}\big[(L-1)(\breve{F}^\ast_{T,L}(x)-\breve{F}^\ast_{L}(x))^2\big]=
   o\big(\frac{\sqrt{\log T}}{\sqrt{T}}\big)+o\big(\frac{L \log T}{T}\big)=o(1)
\end{equation*}
holds uniformly for all $x>0$. This proves (\ref{eqn:arguentofthm2.1-509}).

\textbf{Proof of (\ref{eqn:arguentofthm2.1-508}).} Note that
\begin{align}
 &\int_0^\infty\frac{\sqrt{L-1}(\breve{F}^\ast_{T,L}(x)-\breve{F}^\ast_{L}(x))}{F(x)(1-F(x))}\,\mathrm{d}F(x)
 = \sqrt{L-1}\int_0^\infty \frac{\breve{F}^\ast_{T,L}(x)-\breve{F}^\ast_{L}(x)}{x} \,\mathrm{d} x \nonumber\\
 &\qquad =\frac{1}{\sqrt{L-1}}\sum_{l=1}^{L-1} \int_0^\infty \Big( \mathbb{I}_{(0,x]}\Big(\frac{\phi_{1,T,l}}{\phi_{2,T,l}}\Big) -\mathbb{I}_{(0,x]}\Big(\frac{\chi^2_{1,l}}{\chi^2_{2,l}}\Big)\Big)\frac{1}{x} \,\mathrm{d} x. \label{eqn:arguentofthm2.1-51}
\end{align}
It follows from (\ref{eqn:arguentofthm2.1-50}) that
\begin{align}
 & \Big|\int_0^\infty \Big( \mathbb{I}_{(0,x]}\Big(\frac{\phi_{1,T,l}}{\phi_{2,T,l}}\Big) -\mathbb{I}_{(0,x]}\Big(\frac{\chi^2_{1,l}}{\chi^2_{2,l}}\Big)\Big)\frac{1}{x} \,\mathrm{d} x\Big| =\Big|\log\Big(\frac{\phi_{1,T,l}}{\phi_{2,T,l}}\Big) -\log\Big(\frac{\chi^2_{1,l}}{\chi^2_{2,l}}\Big)\Big) \Big| \nonumber\\
 &\qquad \leq 2 \big|\log\big(1+o_p(1)\big)\big|=o_p\big(1\big) \label{eqn:arguentofthm2.1-52}
\end{align}
holds for all $x>0$ and $l=1,\cdots,L-1$. Combining (\ref{eqn:arguentofthm2.1-51}) and (\ref{eqn:arguentofthm2.1-52}), we obtain (\ref{eqn:arguentofthm2.1-508}).


\begin{thebibliography}{50}
\bibitem[Anderson and Darling(1952)]{AndersonDarling1952}Anderson, T. W. and Darling, D. A. (1952) Asymptotic theory of certain`goodness of fit' criteria based on stochastic processes. \textit{Ann. Math. Statist.} 23, 193--212.
\bibitem[Anderson and Darling(1954)]{AndersonDarling1954}Anderson, T. W. and Darling, D. A. (1954) A test of goodness of fit. \textit{J. Amer. Statist. Assoc.} 49, 765--769.
\bibitem[Billingsley(1995)]{Billingsley1995}Billingsley, P. (1995) \textit{Probability and measure}, 3rd edn. John Wiley \& Sons, New York.
\bibitem[Brillinger(2001)]{Brillinger2001}Brillinger, D.~R. (2001) \textit{Time Series: Data Analysis and Theory}. SIAM, Philadephia.
\bibitem[Brockwell and Davis(1991)]{BrockwellDavis1991}Brockwell, P. and Davis, R.~A. (1991) \textit{Time Series: Theory and Methods}, 2nd edn. New York: Springer.
\bibitem[Dahlhaus(1997)]{Dahlhaus1997}Dahlhaus, R. (1997) Fitting time series models to nonstationary processes. \textit{Ann. Statist.} 25, 1--37.
\bibitem[Dahlhaus et al.(2017)]{Dahlhausetal2017}Dahlhaus, R., Richter, S. and Wu, W. (2017) Towards a general theory for non-linear locally stationary processes. arXiv:1704.02860v3.
\bibitem[DasGupta(2008)]{DasGupta2008}DasGupta, A. (2008) \textit{Asymptotic Theory of Statistics and Probability}. New York: Springer.
\bibitem[Decowski and Li(2015)]{DecowskiLi2015}Decowski, J. and Li, L. (2015) Wavelet-based tests for comparing two time series analysis. \textit{J. Time Ser. Anal.} 36, 189--208.
\bibitem[Dette et al.(2011a)]{Detteetal2011}Dette, H., Kinsvater, T. and Vetter, M. (2011a) Testing non-parametric hypotheses for stationary processes by estimating minimal distances. \textit{J. Time Ser. Anal.} 32, 447--461.
\bibitem[Dette et al.(2011b)]{Dettetal2011b}Dette, H., Preuss, P. and Vetter, M. (2011b) A measure of stationarity in locally stationary processes with applications to testing. \textit{J. Amer. Statist. Assoc.} 106, 1113--1124.
\bibitem[Diggle and Fisher(1991)]{Diggleetal1991}Diggle, P.~J. and Fisher, N.~I. (1991) Nonparametric comparison of cumulative periodograms. \textit{J. R. Stat. Soc. Ser. B Stat. Methodol.} 71, 831--857.
\bibitem[Eichler(2008)]{Eichler2008}Eichler, M. (2008) Testing nonparametric and semiparametric hypotheses in vector stationary process. \textit{J. Multivariate Anal.}, 99, 968--1009.
\bibitem[Faraway et al.(2015)]{Farawayetal2015}Faraway, J., Marsaglia, G., Marsaglia, J. and Baddeley, A. (2015) \textbf{goftest}: Classical goodness-of-fit tests for univariate distributions. URL https://CRAN.R-project.org/package=goftest.
\bibitem[Fuller(1996)]{Fuller1996}Fuller, W.~A. (1996) \textit{Introduction to Statistical Time Series}, 2nd edn. New York: John Wiley.
\bibitem[Guo and Dai(2006)]{GuoDai2006}Guo, W. and Dai, M. (2006) Multivariate time-dependent spectral analysis using Cholesky decomposition. \textit{Statist. Sinica} 16, 825--845.
\bibitem[G\"{o}tze(1991)]{Gotze1991}G\"{o}tze, F. (1991) On the rate of convergence in the multivariate CLT. \textit{Ann. Probab.} 19, 724--739.
\bibitem[Jentsch and Pauly(2015)]{JentschPauly2015}Jentsch, C. and Pauly, M. (2015) Testing equality of spectral densities using randomization techniques. \textit{Bernoulli} 21, 697--739.
\bibitem[Lahiri(2003)]{Lahiri2003}Lahiri, S. N. (2003) A necessary and sufficient condition for asymptotic independence of discrete Fourier transforms under short- and long-range dependence. \textit{Ann. Statist.} 31, 613--641.
\bibitem[Marsaglia and Marsaglia(2004)]{MarsagliaMarsaglia2004}Marsaglia, G. and Marsaglia, J. (2004) Evaluating the Anderson-Darling distribution. \textit{J. Stat. Softw.} 9(2), 1--5.
\bibitem[Shumway and Stoffer(2011)]{ShumwayStoffer2011}Shumway, R.~H. and Stoffer, D.~S. (2011) \textit{Time Series Analysis and its Applications with R Examples}, 3rd edn. New York: Springer.
\bibitem[Szeidl and Zolotarev(1998)]{SzeidlZolotarev1998}Szeidl, L. and Zolotarev, V.~M. (1998) The central limit theorem without the condition of independence. \textit{J. Math. Sci.} 91,3002--3004.
\bibitem[van Bellegem and von Sachs(2008)]{BellegemSachs2008}van Bellegem, S. and von Sachs, R. (2008) Locally adaptive estimation of evoluationary wavelet spectra. \textit{Ann. Statist.} 36, 1879--1924.
\bibitem[Wu(2005)]{Wu2005}Wu, W. (2005) Nonlinear system theory: Another look at dependence. \textit{PNAS} 102, 14150--14154.
\bibitem[Zhang and Tu(2018)]{ZhangTu2018}Zhang, S. and Tu, X. M. (2018) Tests for comparing time-invariant and time-varying spectra based on the Pearson statistic. \textit{J. Time Ser. Anal.} 39, 709--730. \end{thebibliography}
\end{document}